\documentclass[preprint,sort&compress,12pt]{elsarticle}

\usepackage[top=1in, bottom=1in, left=1in, right=1in]{geometry}

\usepackage{amssymb}
\usepackage{latexsym}

\usepackage{url}
\usepackage{xcolor}
\definecolor{newcolor}{rgb}{.8,.349,.1}

\newcommand{\bbE}{\mathbb{E}}
\newcommand{\bbR}{\mathbb{R}}

\newcommand{\cX}{\mathcal{X}}
\newcommand{\cF}{\mathcal{F}}
\newcommand{\cH}{\mathcal{H}}
\newcommand{\cG}{\mathcal{G}}

\newcommand{\cR}{\mathcal{R}}
\newcommand{\cL}{\mathcal{L}}

\newcommand{\diff}{\mathrm{d}}

\newcommand{\phys}{\textrm{P}}
\newcommand{\struct}{\textrm{I}}
\newcommand{\obs}{\textrm{E}}
\newcommand{\local}{\textrm{loc}}
\newcommand{\nonlocal}{\textrm{non-loc}}

\newcommand{\cV}{\mathfrak{V}}

\usepackage{color,soul}
\usepackage{mathtools}

\definecolor{lightblue}{rgb}{.90,.95,1}
\definecolor{darkgreen}{rgb}{0,.5,0.5}
\definecolor{darkred}{rgb}{.7,0,0}

\graphicspath{ {./figs/}}

\usepackage{subfig}
\usepackage{graphicx}
\usepackage{amssymb}
\usepackage{amsthm}
\usepackage{bbm}
\usepackage{bm}
\usepackage{lineno}
\usepackage{url}
\usepackage{listings}
\usepackage[colorlinks=true]{hyperref}

\usepackage{changes}

\definechangesauthor[name={Reviewer 1}, color = blue]{R1}
\definechangesauthor[name={Reviewer 2}, color = red]{R2}
\definechangesauthor[name={All reviewers}, color = brown]{All}
\definechangesauthor[name={Editor}, color = purple]{Editor}
\definechangesauthor[name={Author}, color = olive]{Author}

\DeclareMathOperator*{\argmin}{arg\,min}

\begin{document}

\begin{frontmatter}
\title{Learning About Structural Errors in\\ Models of Complex Dynamical Systems}

\author[1]{Jin-Long Wu\corref{cor1} \fnref{fn1}}
\ead{jinlong.wu@wisc.edu}
\author[2]{Matthew E. Levine \fnref{fn1}}
\author[3]{Tapio Schneider}
\author[3]{Andrew Stuart}
\fntext[fn1]{Jin-Long Wu and Matthew Levine started to work on this research at California Institute of Technology.}
\cortext[cor1]{Corresponding author}

\address[1]{University of Wisconsin-Madison, 1513 University Ave, Madison, WI 53706, USA}
\address[2]{Eric and Wendy Schmidt Center, Broad Institute of MIT and Harvard, Cambridge, USA, 02142}
\address[3]{California Institute of Technology, 1200 E California Blvd, Pasadena, CA 91125, USA}

\begin{abstract}
Complex dynamical systems are notoriously difficult to model because some degrees of freedom (e.g., small scales) may be computationally unresolvable or are incompletely understood, yet they are dynamically important. For example, the small scales of cloud dynamics and droplet formation are crucial for controlling climate, yet are unresolvable in global climate models. Semi-empirical closure models for the effects of unresolved degrees of freedom often exist and encode important domain-specific knowledge. Building on such closure models and correcting them through learning the structural errors can be an effective way of fusing data with domain knowledge. Here we describe a general approach, principles, and algorithms for learning about structural errors. Key to our approach is to include structural error models inside the models of complex systems, for example, in closure models for unresolved scales. The structural errors then map, usually nonlinearly, to observable data. As a result, however, mismatches between model output and data are only indirectly informative about structural errors, due to a lack of labeled pairs of inputs and outputs of structural error models. Additionally, derivatives of the model may not exist or be readily available. We discuss how structural error models can be learned from indirect data with derivative-free Kalman inversion algorithms and variants, how sparsity constraints enforce a ``do no harm'' principle, and various ways of modeling structural errors. We also discuss the merits of using non-local and/or stochastic error models. In addition, we demonstrate how data assimilation techniques can assist the learning about structural errors in non-ergodic systems. The concepts and algorithms are illustrated in two numerical examples based on the Lorenz-96 system and a human glucose-insulin model.
\end{abstract}


\end{frontmatter}

\section{Introduction}
\label{sec:intro}

Numerical simulation is at the heart of modeling, predicting, and understanding dynamical systems that are too complex to be amenable to analytical solution. Complex dynamical systems here extend from molecular dynamics with quantum effects to the planetary scales of weather and climate. The range of dynamically important scales in these systems can be vast, for example, in case of the atmosphere, extending over 13 orders of magnitude from the micrometers of cloud droplets and aerosols to the tens of thousands kilometers of planetary waves. The number of degrees of freedom that would need to be resolved for a faithful simulation of such systems (e.g., $\gtrsim 10^{21}$ for a typical atmospheric boundary layer flow) often exceeds what will be computationally feasible for the foreseeable future \cite{Schneider17a}.

Instead of direct numerical simulation, a variety of approaches has been devised to approximately resolve the most important degrees of freedom in numerical simulations. The degrees of freedom that remain unresolved but, because of nonlinear interactions, are still important for the resolved degrees of freedom are then represented by closure models, which link what is unresolved to what is resolved. The state $X$ of the approximate system evolves according to dynamics of the form
\begin{align}
\label{eq:xdot}
    \dot{X} = f(X; \theta_\phys),
\end{align}
where $f$ may depend on derivatives of the state $X$; hence, the system may represent partial differential equations. The system depends on empirical parameters $\theta_\phys$ that appear in closure models. For example, in large-eddy simulations of turbulent flows, the most energetic ``large eddies'' are explicitly resolved in the dynamics represented by $f$. The effect of the unresolved scales is modeled by subgrid-scale models, such as the classical Smagorinsky model \cite{Smagorinsky63}, which depend on empirical parameters $\theta_\phys$ (e.g., the Smagorinsky coefficient). Similar semi-empirical models are used in many other fields. They encode domain-specific knowledge, and their parameters $\theta_\phys$ need to be calibrated with data.

Data $y$ that are informative about the system come in a variety of forms, such as direct measurements of the time evolution of the state $X$ or more indirect mappings of the state $X$ onto observables, which may, for example, be statistical aggregates of state variables or convolutions of state variables with kernels. Convolutional data arise, for
example, when representing the effect of a state variable such as temperature on the radiative energy fluxes that a satellite measures from space. Generally, we can write that the state maps to observables via an observation operator $\cH$, such that
\begin{align}
    \label{eq:ydag}
    \hat y = \cH[X].
\end{align}
The challenge is that simulated observables $\hat y$ generally are biased estimates of actual data $y$. The actual data $y$ are affected by measurement error, and the simulated data $\hat y$ are affected by structural errors in the approximate dynamical system \eqref{eq:xdot}; both $y$ and $\hat y$ can also be affected by sampling error. For example, while a general feature of turbulence is to enhance mixing of conserved quantities, turbulent mixing is not always diffusive in character. Therefore, diffusive subgrid-scale models such as the Smagorinsky model are not always structurally correct, especially in convective situations with coherent flow structures \cite{holtslag1991eddy}. This can lead to biases that, for example, adversely affect the calibration of model parameters $\theta_\phys$.

The purpose of this paper is to summarize principles of, and algorithms for, learning about structural error models that correct semi-empirical closure models. Wholesale replacement of semi-empirical closure models with neural networks and other deep learning approaches promises to overcome the structural strictures of existing closure models through more expressive models; it has recently received much attention \cite{ling2016reynolds,weatheritt2016novel,Rasp18a,maulik2019subgrid,duraisamy2019turbulence,zhao2020rans,schmelzer2020discovery,Brunton20a}. However, existing semi-empirical closure models encode valuable domain-specific knowledge. Learning flexible corrections to these models is often less data hungry, more interpretable, and  potentially more generalizable than replacing them wholesale.

What follows is a distillation of experiences we gained in studying various complex dynamical systems. Our goal is to provide guidelines and algorithms that can lead to a broad-purpose computational framework for systematically learning about model error in dynamical systems. We focus on two important parts of error modeling: (i) how to construct an error model, and (ii) how to calibrate an error model.

Our approach to constructing error models builds upon but goes beyond the classical work of Kennedy and O'Hagan~\cite{kennedy2001bayesian}, who accounted for model error through an external bias correction term $\delta(X; \theta_\obs)$, parameterized by parameters $\theta_\obs$ and inserted at the boundary between output from a computer model $\hat y = \mathcal{G}(\theta_\phys)$ and data $y$:
\begin{equation}
y = \hat y + \delta(X; \theta_\obs) + \eta.
\end{equation}
Here, $\mathcal{G}(\theta_\phys) = \mathcal{H}[X(\theta_\phys)]$ corresponds to solving \eqref{eq:xdot} for the time series of the state $X$, which depends parametrically on $\theta_\phys$, and then applying the observation operator \eqref{eq:ydag}; hence, $\mathcal{G}$ is a mapping from the space of model parameters $\theta_\phys$ to the space of observations $y$. The noise $\eta$ represents additional (e.g., observation) errors, assumed to have zero mean. In this approach, the model parameters $\theta_\phys$ remain fixed (i.e., a property of $\mathcal{G}$) while parameters $\theta_\obs$ in the error model $\delta$ are tuned such that the residual $y - \hat y$ has a small magnitude and zero mean. This approach of externalizing model error for bias correction has been applied and further expanded in many subsequent papers \cite[e.g.,][]{higdon2004combining,wang2009bayesian,brynjarsdottir2014learning,trehan2017error,calvetti2018iterative}. A key advantage of the external model error approach is that the model producing $\hat y$ can be treated as a black box, which facilitates use of this approach across different domains. State variables $X$ and residuals $y - \hat y$ form input-output pairs from which the error model $\delta(X; \theta_\obs)$ can be learned, for example, with supervised learning\footnote{Supervised learning refers to regression or interpolation of data.}  approaches, usually in an offline setting separate from learning about the model parameters $\theta_\phys$. 

However, the external model error approach has several drawbacks:
\begin{itemize}
    \item It is difficult to incorporate physical (or other process-based) knowledge or constraints (e.g., conservation laws) in the error model $\delta(X; \theta_\obs)$ \cite{brynjarsdottir2014learning}.
    \item It cannot improve predictions for quantities other than the observations $y$ on which the error model $\delta(X; \theta_\obs)$ has been trained.
    \item It leads to interpretability challenges because $\delta(X; \theta_\obs)$ is a catch-all error term that typically represents the sum total of errors made in several, and often disparate, closure models.
\end{itemize}

\begin{figure}[!htbp]
  \centering
  \includegraphics[width=0.8\textwidth]{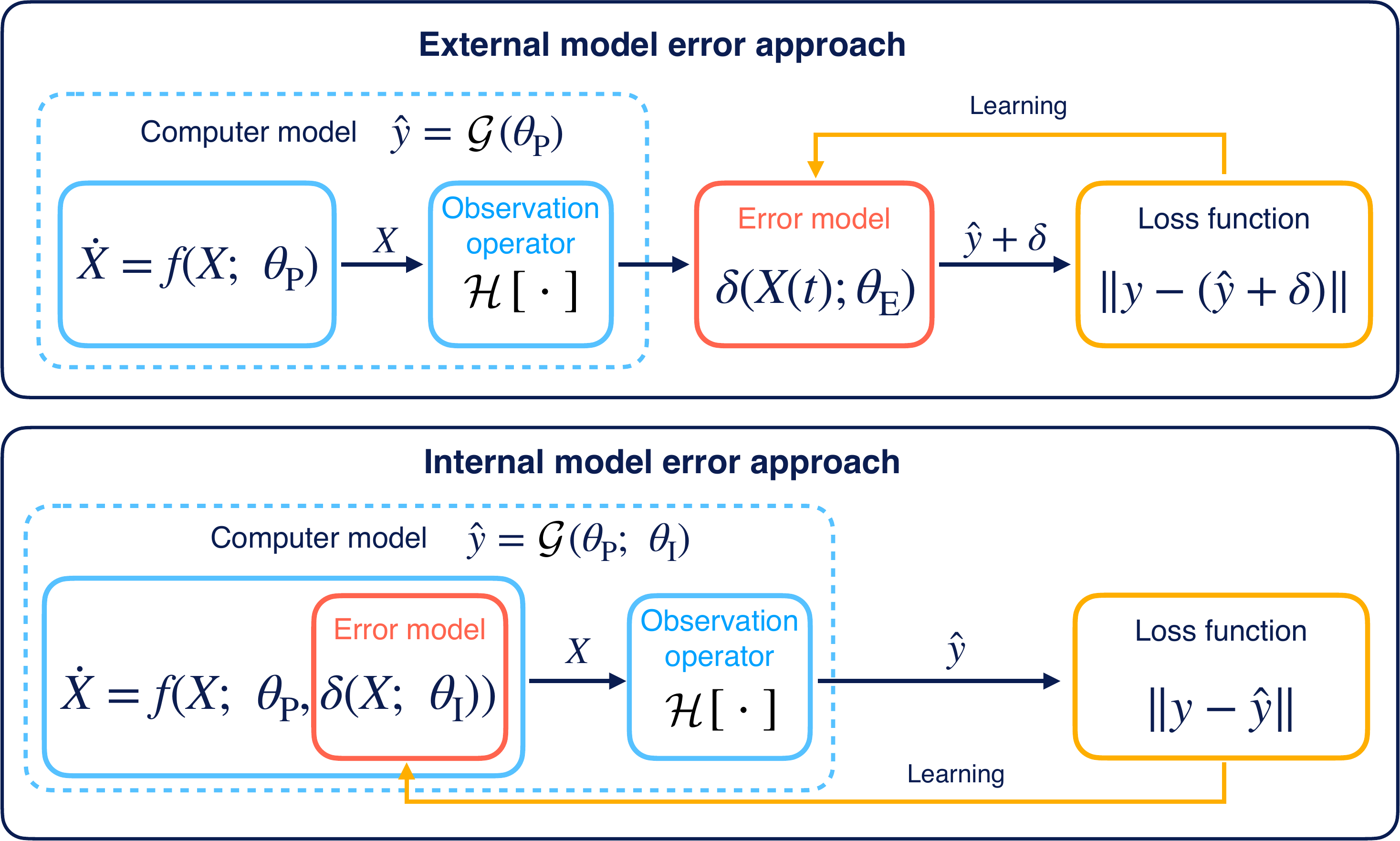}
    \caption{External and internal approaches to modeling structural errors in complex dynamical systems.}
  \label{fig:graphic_abstract}
\end{figure}

To address these drawbacks, a few researchers have started to explore an approach that internalizes model error \cite{strong2014model,he2016numerical,sargsyan2019embedded}. Such an internal model error approach embeds $L$ error models $\delta_l(X;\  \theta_\struct^{(l)})$ ($l=1, \dots, L$) within the dynamical system, at the places (e.g., in closure models) where the errors are actually made.
Let their collection be written as
\begin{equation}
\label{eq:deltaL}
\delta(X;\ \theta_\struct) := \Big\{\delta_l\big(X \ ;\  \theta_\struct^{(l)}\big)\Big\}_{l=1}^L
\end{equation}
so that we can write for the overall system
\begin{equation}
\label{eq:xdot_struct}
\dot{X} = f\Big(X;\ \theta_\phys,
\delta(X;\ \theta_\struct)\Big).
\end{equation}
The error models internal to the dynamical system are chosen so that the error-corrected computer model $\hat{y}=\mathcal{G}(\theta_\phys;\ \theta_\struct) = \mathcal{H}[X(\theta_\phys;\ \theta_\struct)]$ provides unbiased estimates of the data $y$:
\begin{equation}
\label{eq:inv_problem}
y = \mathcal{G}(\theta_\phys;\ \theta_\struct) + \eta,
\end{equation}
where the additional errors $\eta$ are still assumed to have zero mean. Figure~\ref{fig:graphic_abstract} illustrates and contrasts the external and internal approaches to modeling structural errors. 

Such an approach has found applications, for example, in turbulence modeling \cite{emory2011modeling,oliver2011bayesian,cheung2011bayesian,xiao2016quantifying,pernot2017critical,wang2017physics,wu2018physics}. By incorporating the structural error models $\delta(X;\ \theta_\struct)$ inside the dynamical system, the error models can in principle lead to improved predictions even of quantities that were not used to train the error models. The error models can be learned alongside the model parameters $\theta_\phys$ in an online setting. They also are more amenable to interpretation because they are included in the places where errors are actually made. A potential downside of internalizing model error is that the effects of the structural errors $\delta$ map onto data $y$ only through the resolved state $X$ of the dynamical system, and residuals $y - \hat y$ are generally not directly informative about how the structural errors $\delta(X;\ \theta_\struct)$ depend on state variables $X$; thus, learning about structural errors $\delta(X;\ \theta_\struct)$ can generally not be accomplished with direct supervised learning approaches. Instead, residuals $y - \hat y$ only provide indirect information about structural errors $\delta(X;\ \theta_\struct)$. Additionally, if derivatives of the dynamical system $f$ with respect to parameters are not easily available, or if the dynamical system is not differentiable, gradient-based methods for learning about the model errors $\delta(X;\ \theta_\struct)$ are difficult or impossible to use. 

Here we show various ways of constructing models for structural errors and demonstrate how one can learn about the structural errors from direct or indirect data in the absence of derivatives of the dynamical system. As error models, we will consider:
\begin{itemize}
    \item Gaussian processes, as in Kennedy and O'Hagan~\cite{kennedy2001bayesian};
    \item Models assembled from dictionaries of terms (e.g., involving differential operators), as in the data-driven discovery of partial differential equations \cite{schaeffer2013sparse,rudy2017data,schaeffer2017learning,zanna2020data,Silva20b,Lemos22t};
    \item Neural networks, for their expressivity \cite{ling2016reynolds,brenowitz2018prognostic,Rasp18a,maulik2019subgrid,zanna2021deep,Bretherton22i};
    \item Stochastic models, because without a clear scale separation between resolved and unresolved degrees of freedom, theoretical analysis suggests that closure models generally should be stochastic \cite[e.g.,][]{zwanzig2001nonequilibrium,Lucarini14a,franzke2015stochastic,palmer2019stochastic};
    \item Non-local models, because structural errors may be non-local in space \cite{zhou2021learning,zhuang2021learned,du2012analysis,d2016identification,pang2020npinns}, in time \cite{ma2018model,wang2020recurrent,lin2021data}, or in both~\cite{charalampopoulos2021machine}.
\end{itemize}

We will discuss how to learn about such error models both from direct and indirect data. Supervised learning of various types of error models $\delta_l$ from direct data has been performed in a number of settings, for example, to discover terms in differential equations or neural network closure models from residual time tendencies that give direct information about the error model that is sought \cite[e.g.,][]{brunton2016discovering,wang2017physics,wu2018physics,levine2022framework}. However, data directly informative about error models, such as high-resolution time tendencies, are not always available. When training an error model on time tendencies, it can also be difficult to ensure both stability of the dynamical system including the error model and satisfaction of physical constraints (e.g., energy conservation) \cite[e.g.,][]{Bretherton22i}. Training an error model with indirect data, in an inverse problem rather than supervised learning setting \cite[e.g.,][]{xiao2016quantifying,schneider2020learning,zhang2022ensemble}, can be advantageous in those circumstances. We demonstrate how it can be accomplished.

Our main contributions are:
\begin{itemize}
    \item Summarizing the internal model error approach and comparing and contrasting it with the external model error approach.
    \item Presenting a comprehensive study of the internal model error approach for various model forms, including dictionary-based models, Gaussian processes, and neural networks, including non-local and stochastic generalizations.
    \item Demonstrating various methods for calibrating the internal model error, including (i) learning from direct data, (ii) learning from indirect data (e.g., moments of the time series for ergodic systems) with ensemble Kalman inversion (EKI), and (iii) learning from the time series of non-ergodic systems and using EKI for data assimilation (DA). 
    \item Illustrating the importance of constraints (i.e., sparsity and known physics) for improved generalization of the calibrated internal model error.
\end{itemize}

The principles and algorithms we will discuss are broad purpose and applicable across a range of domains and model complexities. To illustrate their properties in a relatively simple setting, we use two versions of the Lorenz 96 \cite{lorenz1996predictability} dynamical system: the basic version of the model and its multiscale generalization
\cite{fatkullin2004computational}. Section~\ref{sec:calibration} discusses the calibration of internal error models with direct or indirect data and the enforcement of constraints such as conservation properties. Section~\ref{sec:construction} introduces various ways of constructing error models. Section~\ref{sec:examples} introduces the two Lorenz 96 systems and
then proceeds to present various concepts and methods through numerical results for these systems. The Lorenz 96 systems serve as illustrative examples of ergodic systems, for which we avoid trajectory matching and learn about unknown parameters from time-averaged statistics, which approximate the expectation with respect to the invariant measure. Section~\ref{sec:examplesGI} is organized similarly to the previous section, but deals with non-ergodic systems for which trajectory matching is needed to learn about the underlying unknown parameters; as an example, we study a model of the human glucose-insulin system. In Section~\ref{sec:conclusions}, we state our conclusions.

\section{Calibrating Error Models}
\label{sec:calibration}
We first summarize several important aspects of calibrating internal error models, including (i) direct or indirect data, (ii) gradient-based or derivative-free optimization, and (iii) enforcing constraints (e.g., sparsity or physical laws). We discuss these aspects with a generic internal error model $\delta(X;\theta_\struct)$, which may represent any one of the error model types we will introduce later.

\subsection{Data Availability}
\subsubsection{Direct Data}

Direct data to calibrate $\delta$ are defined as ``labeled'' input-output pairs $\{X(t_i), \delta(X(t_i))\}_{i=1}^N$, where $i$ denotes a time index. Consider the additive error model $$\dot{X} = f(X; \theta_\phys) + \delta(X;\theta_\struct)$$ as an example. A fine temporal resolution of $X(t)$ is usually needed to approximate $\dot{X}-f(X; \theta_\phys)$ and obtain estimates of the error $ \delta(X;\theta_\struct)$ as a residual. With this method, it becomes challenging to obtain reliable direct data when the trajectories $\dot X$ are noisy, for example, when the dynamical system is chaotic~\cite{brunton2016discovering}. Furthermore it may not be possible to observe
the entirety of $X$. This may be handled as a missing data problem \cite{little2019statistical}, and could be handled by joint parameter-state estimation
for example using data assimilation; see \cite{bocquet2019data,chen2022autodifferentiable}.

An additional complication with  using direct data is ensuring the stability of the dynamical system with the calibrated error model $\delta(\cdot)$. Although with direct data, we can get more control of the accuracy of the error model itself, the calibrated error model often leads to unstable simulations of the dynamical system with the error model \cite{beck2019deep,brenowitz2020interpreting}. There are several ways to mitigate the instability introduced by the error model, e.g., adopting a structure that ensures physical constraints \cite{yuval2021use}, enforcing physical constraints \cite{beucler2021enforcing}, ensuring stability by bounding the eigenvalues of the linearized operator, and limiting the Lipschitz constant of $\delta(\cdot)$ \cite{shi2019neural}. However, a systematic approach to ensure stability is lacking.

\subsubsection{Indirect Data}
Instead of assuming access to direct data $\{X, \delta(X)\}$, the error model can also be calibrated with indirect data by solving an inverse problem associated with \eqref{eq:inv_problem} (i.e., solve for the most likely parameters $\theta_\phys, \theta_\struct$ given the model $\mathcal{G}$ and data $y$).
Using indirect data involves simulating the dynamical system with the error model as in \eqref{eq:xdot_struct}; therefore, the calibration procedure with indirect data favors error models that lead to stable simulations, an important advantage over the direct methods. Typical examples of problems giving rise to indirect data include time-series of $X$ for which the resolution is not fine enough to extract direct data for calibration \cite{boys2008bayesian}, time-averaged statistics of $X$ \cite{schneider2017earth} or dynamical systems that are partially observed \cite{pokern2009parameter}. More generally, indirect data can also be interpreted as constraints on $X$, and thus physical constraints can be enforced via augmenting indirect data.

\subsection{Methods of Calibration}
Using direct data $\{X, \delta(X)\}$ for calibration leads to a regression problem, which can be solved with standard methods for a given parameterization of the error model (e.g., least squares fit for dictionary learning, gradient descent methods for neural network). By contrast, using indirect data for calibration leads to an inverse problem associated with Eq.~\eqref{eq:inv_problem}. Indirect methods can be linked to direct methods by framing the inverse problem as a missing data problem \cite{little2019statistical} 
and alternating between updating the missing
data and updating the calibration parameters using learned direct data, for
example, using the expectation-maximization (EM) algorithm \cite{meng1997algorithm}. However, in this section we focus on the calibration in the inverse problem setting; we discuss gradient-based and derivative-free optimization methods and how to enforce constraints.

\subsubsection{Gradient-based or Derivative-free Optimization}
Eq.~\eqref{eq:inv_problem} defines a forward problem in which $\mathcal{G}, \theta_\phys, \theta_\struct$ and noise $\eta$ can be used to generate simulated data.
The associated inverse problem involves identifying the most likely parameters  $\theta_\phys, \theta_\struct$ for $\mathcal{G}$, conditioned on observed data $y$. To formalize this, we first define a loss function
\begin{equation}
\label{eq:loss}
\begin{aligned}
    \cL(\theta)=&\frac12\bigl|y-\cG\bigl(\theta_\phys;\ \theta_\struct\bigr)\bigr|^2_{\Sigma},
\end{aligned}
\end{equation}
where $\Sigma$ denotes the covariance of the zero-mean noise $\eta$, which is assumed to be independent of $\theta$.\footnote{By $| \cdot |_B$, we denote the covariance-weighted norm defined by $|v|_B = v^* B^{-1} v$ for any positive definite $B$.} 
The inverse problem 
$$\theta^*=\argmin_\theta \cL(\theta)$$ can be solved by gradient descent methods once the gradient
\begin{equation}
    \frac{d \cL}{d \theta}=\frac{d \cG^T}{d \theta}\Sigma^{-1}(y-\cG)
\end{equation}
is calculated, where $\theta=[\theta_\struct,\theta_\phys]$ collects all parameters. In practice, the action of the gradient $d \cG^T / d \theta$ is often evaluated via adjoint methods for efficiency. Although the gradient-based optimization is usually more efficient when $\cG$ is differentiable, the evaluation of $\cG$ can be noisy (e.g., when using finite-time averages to approximate infinite-time averaged data \cite{dunbar2022ensemble}) or stochastic (e.g., when using stochastic processes to construct the error model). In these settings, gradient-based optimization may no longer be suitable, and derivative-free optimization becomes necessary. In this
paper we focus on Kalman-based derivative-free optimization for solving the inverse problem; \ref{sec:eki} briefly reviews a specific easily implementable form of ensemble Kalman inversion (EKI), to illustrate how the methodology works, and gives pointers to
the broader literature in the field.

It is worth noting that EKI requires ensemble simulations, which may be less attractive when the system is not chaotic and gradient information is available. Even for chaotic systems, matching short trajectories can be achieved by gradient-based optimization~\cite{chen2023operator}, without the need to run ensemble simulations. For large datasets, the computational cost of EKI increases, e.g., because of the cost of the matrix inversions in the EKI formula. However, it is possible to make use of the sparse structure of the covariance matrix in data space and thus reduce the computational cost of the matrix inversion~\cite{bollhofer2019large}. For a large dataset and a large number of parameters, a relatively small ensemble size may not suffice to approximate the covariance matrix; in that case, localization techniques~\cite{greybush2011balance,tong2023localized,liu2023dropout} can be used to avoid spurious correlations due to the small ensemble size. In practice, localization techniques often lead to sparse covariance matrices, which allows users to employ standard efficient algorithms for the inverse and the multiplication of those covariance matrices in EKI.

\subsubsection{Enforcing Constraints}
\label{ssec:constraints}

There are various types of constraints that can be enforced when calibrating an error model. Two common constraints are sparsity constraints and physical constraints (e.g., conservation laws). Here we present the general concept of enforcing these two types of constraints in calibration, and \ref{sec:eki} presents more details about using EKI to solve the corresponding constrained optimization problems.

Sparsity is important to impose on error models to avoid adding parameters to an error model that have no or little measurable impact on the output $\hat y$ of the host model; that is, the first principle in modeling structural errors should be to do no harm that would result from unnecessary error model complexity. To impose sparsity on the solution of $\theta_\struct$, we aim to solve the optimization problem~\cite{schneider2020ensemble}
\begin{equation}
\label{eq:optim-EKI-lasso}
\begin{aligned}
    \cL(\theta;\lambda):=&\frac12\bigl|y-\cG\bigl(\theta_\phys;\ \theta_\struct\bigr)\bigr|^2_{\Sigma}+\lambda|\theta_\struct|_{\ell_0},\\
    \theta^*=&\argmin_{\theta \in \cV} \cL(\theta;\lambda),
\end{aligned}
\end{equation}
where $\cV=\{\theta: |\theta_\struct|_{\ell_1} \le \gamma\}$. The regularization parameters $\gamma$ and $\lambda$ can be determined via cross-validation. In practice, adding the $\ell_0$ constraint is achieved by thresholding the results from $\ell_1$-constrained optimization. The detailed algorithm was proposed in~\cite{schneider2020ensemble} and is summarized in \ref{sec:eki}.

In many applications, it is important to find error models that satisfy physical constraints, such as energy conservation. To impose physical constraints on the solution of $\theta_\struct$ from EKI, we first generalize the constraint as:
\begin{equation}
\label{eq:cset1}
\cV=\{\theta: \cR(\theta_\struct) \le \gamma\}.
\end{equation}
Here, $\cR$ can be interpreted as a function that evaluates the residuals of certain physical constraints (typically, by solving Eq~\eqref{eq:xdot_struct}). The constraint function $\cR$ can be nonlinear with respect to $\theta_\struct$. Taking the additive error model $\dot{X}=f(X)+\delta(X;\theta_\struct)$ as an example, the function $\cR$ corresponding to the energy conservation constraint can be written explicitly as
\begin{equation}
\label{eq:constraint_1}
\cR(\theta_\struct) = \Big|\int_0^T \big(\langle \delta(X(t);\theta_\struct), X(t)\rangle\big) dt \Big|,
\end{equation}
which constrains the total energy introduced into the system during the time interval $[0,T]$. Alternatively, a stronger constraint can be formulated as
\begin{equation}
\label{eq:constraint_2}
\cR(\theta_\struct) = \int_0^T \big|\langle\delta(X(t);\theta_\struct), X(t)\rangle\big| dt,
\end{equation}
which constrains the additional energy introduced into the system at every infinitesimal time step within the time interval $[0,T]$. The notation $\langle \cdot, \cdot \rangle$ denotes the inner product. Both forms of constraint in \eqref{eq:constraint_1} and \eqref{eq:constraint_2} can be implemented by using augmented observations, i.e., including the accumulated violation of the energy constraint as a additional piece of observation data whose true mean value is zero.

\subsubsection{Accounting for Initial Conditions}
\label{ssec:ics}

Evaluation of $\cG$ typically requires an initial condition, $X(0)$, when solving Eq.~\eqref{eq:xdot_struct}.
When we observe ergodic systems over sufficiently long time scales, we can treat $\cG$ as a stochastic map over a distribution of initial conditions \cite{cleary2021calibrate}.
However, when we observe systems over shorter time scales (or if they are non-ergodic), it is often necessary to identify initial conditions that generated the observations. We highlight two basic approaches: (i) append $X(0)$ to the vector of learned parameters for $\cG$, and (ii) implicitly identify $X(0)$ by applying data assimilation techniques. To achieve (i), we can minimize:
\begin{equation}
\label{eq:ic_joint}
    \cL\bigl(\theta, X(0)\bigr) = \frac12\bigl|y-\cG\bigl(\theta_\phys;\ \theta_\struct, X(0) \bigr)\bigr|^2_{\Sigma}.
\end{equation}
To achieve (ii), we can minimize:
\begin{equation}
\label{eq:ic_DA}
\begin{aligned}
    \cL(\theta) = & \frac12\bigl|y-\cG\bigl(\theta_\phys;\ \theta_\struct, X(0) \bigr)\bigr|^2_{\Sigma},\\
    \textrm{s.t.} \ X(0) =& \mathcal{DA}(\theta, y),
\end{aligned}
\end{equation}
where $\mathcal{DA}$ represents a data assimilation algorithm that performs state estimation given observations $y$ and assumed parameters $\theta$.
While Eqs.~\eqref{eq:ic_joint} and \eqref{eq:ic_DA} appear quite similar, the utilization of efficient state-estimation algorithms in \eqref{eq:ic_DA} can help cope with sensitivities to initial conditions that may cause \eqref{eq:ic_joint} to fail.
Recent work has explored different variants of \eqref{eq:ic_DA}, often using gradient-based optimization approaches and sequential state-estimation schemes \cite{levine2022framework,ribera2022model,brajard2021combining,chen2022autodifferentiable}.

\section{Constructing Error Models}
\label{sec:construction}
We highlight three different approaches to representing structural errors: dictionary learning, Gaussian processes, and neural networks; however, other representations of structural error can also be considered
within the overarching framework proposed here. For these three approaches, existing work mainly focuses on constructing deterministic error models that are locally dependent on state variables; however, the approaches can all be extended to the construction of stochastic error models or can be made non-locally dependent on state variables as described in Sections \ref{ssec:stochastic} and \ref{ssec:nonlocal}.
For simplicity, we define the error models for the whole collection of structural errors $\delta(X, \theta_\struct)$; however, we can also define and learn them independently for each component of the structural error model $\delta_l(X, \theta_\struct^l)$ for $l=1, \dots, L$.

\subsection{Dictionary Learning}
\label{sec:dictionary}

If a set of candidate terms in error models is known or can be approximated, an error model can be constructed via learning from a dictionary of $J$ candidate terms,
\begin{equation}
\label{eq:dl}
\delta(X; \theta_\struct) = \sum_{j=1}^J \alpha_{j} \phi_j(X; \beta_{j}),
\end{equation}
where $\theta_\struct = \{\alpha, \beta\}$ and $\phi_j(X; \beta_{j})$ denote user-specified, parametric basis functions that can, for example, include differential operators \cite{brunton2016discovering,rudy2017data,schaeffer2013sparse,schaeffer2017learning}.
In practice, it is difficult to know all suitable basis functions a priori, and thus it is common to include redundant basis functions in the dictionary. Basis functions can then be pruned based on data by imposing sparsity constraints on the coefficients $\alpha_j$. Such sparsity constraints have proven to be beneficial in the construction of data-driven models of dynamical systems \cite{brunton2016discovering,rudy2017data,schaeffer2013sparse,schaeffer2017learning,schneider2020ensemble}. They are also commonly used in compressed sensing \cite{donoho2006compressed},  where dictionary learning has been widely used.

An advantage of using dictionary learning is the potential interpretability of the constructed error model, arising because the error model is a linear combination of user-specified and hence interpretable basis functions. On the other hand, this approach can be overly restrictive when the dictionary of basis functions $\{\phi_j\}$ is misspecified, resulting in an insufficiently expressive error model.

\subsection{Gaussian Processes}
\label{sec:gaussian}

Another option of constructing an error model is via Gaussian processes (GPs)~\cite{williams2006gaussian},\footnote{Here we use
only the mean of the GP, and the methodology is simply a form of
data-adapted regression; we are not using the uncertainty
quantification that comes with GPs.}
\begin{equation}
\delta(X;\theta_\struct) \sim \mathcal{GP}\left(m,\mathcal{K}\right),
\end{equation}
where $m: \mathcal{X} \mapsto \mathbb{R}$ denotes the mean of $\delta$, $\mathcal{K}: \mathcal{X} \times \mathcal{X} \mapsto \mathbb{R}$ represents a kernel, and $\mathcal{X}$ is the input space of the structural error model. Given data at $J$ different points $X^{(j)}$ for $j=1,2,...,J$, the mean of the error model can be written as a linear combination of basis functions,
\begin{equation}
\label{eq:gp}
    m(X)=\sum_j \alpha_{j} \mathcal{K}(X^{(j)}, X;\psi),
\end{equation}
where $\psi$ denotes the hyper-parameters of the kernel $\mathcal{K}$. Therefore, the parameters that characterize the error model become $\theta_\struct=\{\alpha,\psi\}$ if the mean of a GP is used to represent the model error term $\delta$.  The GP approach requires the choice of a kernel $\mathcal{K}$, which then determines the kernel functions $\mathcal{K}(X^{(j)}, \cdot)$ in Eq.~\eqref{eq:gp}.
This may appear restrictive, but the hyper-parameters
of $\mathcal{K}$ are learned from the data; thus, the set of functions in
which the solution is sought is data-adapted. This confers a potential advantage over
dictionary learning, in particular for problems lacking in strong prior knowledge about the functional form of the model $\delta(X; \theta_\struct)$ to be learned. In the case of indirect data, the locations $X^{(j)}$ must also be chosen a priori (or learnt as additional parameters).

Because of the similar forms of Eqs.~\eqref{eq:gp} and \eqref{eq:dl}, the GP shares similar shortcomings as dictionary learning when the kernel $\mathcal{K}$ is misspecified, even in the presence of hyper-parameter learning. In practice, a more sophisticated kernel $\mathcal{K}=\sum\limits_i \mathcal{K}_{i}$ is often constructed from some basic kernels $\mathcal{K}_{i}$ \cite{hamzi2021learning,darcy2021learning,lee2022learning}. If a redundant set of basic kernels is used, sparsity constraints can be imposed in a similar way as in dictionary learning to prune the kernel set. A further limitation of using GPs is the computational cost, which grows exponentially with the dimension of $\mathcal{X}$. This pathology can be ameliorated by representing the GP as a linear combination of random Fourier features \cite{rahimiRandomFeaturesLargeScale2008}, which allows us to recast a GP as a dictionary-based approach in which the bases $\phi_j$ are drawn randomly from a special distribution known to reproduce a kernel of interest.

\subsection{Neural Networks}
\label{sec:neural}

Compared to dictionary learning, neural networks are more expressive, and they are more scalable than GPs, as the latter suffer from the curse of dimensionality if the model has high-dimensional input. Neural networks can also be used to construct an error model,
\begin{equation}
\label{eq:nn}
\delta(X;\theta_\struct) = \mathcal{NN}(X;\theta_\struct),
\end{equation}
where $\mathcal{NN}$ denotes a neural network and $\theta_\struct$ the coefficients (biases and weights) of the neural network. While neural networks are expressive and scalable, it is more difficult to enforce stability of a dynamical system with a neural network error model~\cite{brenowitz2020interpreting}. This is mainly because the nonlinearity introduced by a neural network is often more difficult to analyze compared with dictionary learning, for which we explicitly specify basis functions and thus can avoid using those that lead to instability; it is also more difficult to
analyze than GP based learning because the latter is easier to interpret,
as the kernels are hand-picked and then tuned to data.
In Section \ref{ssec:constraints}, we discuss a general approach to enhancing stability by enforcing energy constraints in the context of learning from indirect data.

\subsection{Stochastic Extension}
\label{ssec:stochastic}

In the preceding sections, we briefly summarized commonly used tools for constructing error models. All of those models were deterministic, with fixed parameters $\theta_\struct$. To quantify uncertainties, we can take a Bayesian perspective, view the unknown parameters as random variables, and infer the distributions of those parameters given the data. We can then propagate the uncertainties of those parameters to the simulated state $X$ and predicted observations $\hat y$ via Monte Carlo simulations. Although this is a standard approach to quantifying uncertainties, it cannot directly account for the impact of neglected information of unresolved scales upon the resolved state $X$. The randomness of the unresolved state can have an order one impact upon $X$; this issue is particularly prevalent in applications without a clear scale separation, such as turbulence, but can also happen in scale-separated problems. In such a scenario, directly modeling this impact as randomness in the resolved state becomes more appropriate, and it can be achieved by further adding a stochastic process to the deterministic error model:
\begin{equation}
\label{eq:stochastic}
\delta\Big(X; \theta_\struct\Big) \diff t= \delta_\mathrm{det}(X; \theta_\mathrm{det}) \diff t +\sqrt{\sigma^2(X;\theta_\mathrm{ran})}\diff W,
\end{equation}
where $\mathrm{det}$ indicates a deterministic model, $W$ denotes the Wiener process, and the overall unknown parameters are defined as $\theta_\struct = \{\theta_\mathrm{det},\theta_\mathrm{ran}\}$. In practice, the above formulation can be further generalized by using stochastic processes (e.g., with desired temporal correlations) other than the Wiener process. 

Fitting stochastic models to time-series data has been explored in some previous 
works~\cite{wilkinson2009stochastic,roberts2001inference,roberts2002langevin,papaspiliopoulos2013data}; a common problem when applying these methods
is the inconsistency between data and the incremental structure of the Gaussian noise
driving the model as the time step is approaching zero~\cite{zhang2005tale,pavliotis2007parameter,papavasiliou2009maximum,boninsegna2018sparse}. A common practice to address this issue is the multi-scale use of data, e.g., via subsampling~\cite{zhang2006efficient,pokern2009parameter,papaspiliopoulos2012nonparametric,pavliotis2012parameter,batz2018approximate,abdulle2020drift}. Some previous works also explored Kramers–Moyal averaging with finite sampling rate correction~\cite{honisch2011estimation,lade2009finite,callaham2021nonlinear}. On the other hand, fitting a discretized version of stochastic processes to time-series data has been explored using autoregressive models~\cite{Neumaier01a, Schneider01c, arnold2013stochastic,lu2020data}. For some dynamical systems, the unresolved state has conditional (with respect to the resolved state) Gaussian statistics~\cite{chen2016filtering,chen2018conditional}, and then fitting the stochastic models can be achieved using analytically derived likelihoods.

In the absence of the whole trajectories of time series, some recent works started to explore fitting stochastic models to statistics of time-series data~\cite{krumscheid2013semiparametric,krumscheid2015data,kalliadasis2015new,schneider2020learning}. Using time-averaged data to estimate linear SDEs has been studied for decades to account for climate variability~\cite{hasselmann1976stochastic,frankignoul1977stochastic,penland1993prediction,Schneider99a}, and extension to nonlinear SDEs was discussed in~\cite{hasselmann1988pips}.

\subsection{Representing Non-local Effects}
\label{ssec:nonlocal}

\paragraph{Spatial Non-locality} The states $X(t)$ for approximate models and their structural corrections typically consider $X(t)$ as a discretized spatial field. Most traditional closure models are formed locally; that is, they rely on the assumption of local dependence on $X(t,r)$, where $X(t,\cdot): \mathbb{R}^p \mapsto \mathbb{R}$ is a spatial field, and $r \in \mathbb{R}^p$ represents the spatial coordinate. 
For some applications, it is useful to consider non-local effect in the error model. Indeed, our formulations of the approximate physical model in \eqref{eq:xdot} and models for structural error in Sections~\ref{sec:dictionary} to \ref{sec:neural} are well-specified for scalar (local, component-wise) or vector-valued (non-local) $X$. Moreover, we note that non-local functions of the state $X(t)$ are best conceptualized as function-valued \emph{operators}---while they take as inputs a vector of neighboring coordinates from $X(t)$, this vector represents a discretized spatial function. Thus, when designing spatially non-local closures, it is often sensible to build them to be consistent across different spatial discretizations. 

In the case of neural networks, we can build spatially non-local closures with convolutional neural networks (CNNs); the Fourier neural operator (FNO) \cite{li2020fourier} or deep operator network (DeepONet) \cite{lu2021learning} provide an extension to an operator limit.
Similarly, GPs and random feature methods (a dictionary-based formulation of GPs) can be designed with spatially non-local vectorized inputs from $X(t)$. Recent theoretical work has  also allowed these basic methods to be taken to a continuous operator limit \cite{nelsenrandom2020,chen2021operatorgp}.

As an emerging topic in the context of data-driven modeling, some recent works have explored non-local diffusion~\cite{du2012analysis,d2016identification,du2019nonlocal,pang2020npinns} and spatially non-local modeling~\cite{zhou2021learning,zhuang2021learned,charalampopoulos2021machine}.
In this work, we capture the spatially non-local dependence on $X$ via a data-driven convolution kernel:
\begin{equation}
\label{eq:non_local}
\delta\Big(X(t,r); \theta_\struct\Big)=\int_{r^\prime \in \Omega} \delta_\local(X(t,r^\prime);\theta_\local) \mathcal{C}(r-r^\prime;\theta_\nonlocal) \, dr^\prime
\end{equation}
where $\Omega \subset \mathbb{R}^p$ represents a subset of $\mathbb{R}^p$ that contains $r$, and $\mathcal{C}: \mathbb{R}^p \mapsto \mathbb{R}$ denotes a convolution kernel with hyper-parameters $\theta_\nonlocal$. The overall parameterization is defined by $\theta_\struct = \{\theta_\local,\theta_\nonlocal\}$, such that the unknown parameters in the local error model $\delta_\local$ and the convolutional kernel $\mathcal{C}$ can be jointly estimated.

Note that hyper-parameters can be made state-dependent, so that $\theta_\nonlocal(X(t,r);\kappa)$; in this case, the additional unknowns $\kappa$ can be learned,  appending them to $\theta_\struct$. Similarly, learning a nonlinear integral kernel has been discussed in~\cite{kovachki2021neural} and shown to be a continuous generalization of the transformer architecture~\cite{vaswani2017attention}.

The form of non-local closure in \eqref{eq:non_local} draws inspiration from a series of works about non-local modeling~\cite{du2019nonlocal}, in which $\delta_\local$ corresponds to a local Laplace operator. Some mathematical foundations of non-local operators and calculus were summarized in~\cite{du2019nonlocal}, and the connection to fractional differential operators was illustrated in~\cite{bucur2016nonlocal}.

\paragraph{Temporal Non-locality} 

Non-locality in time (memory) is also important. Generically, any form of variable elimination
or coarse-graining results in memory effects which require, at each current point in time,
integration of the entire time-history from the initial condition up to the current
time \cite{zwanzig2001nonequilibrium}. Such memory effects are undesirable as they lead to computational algorithms that scale poorly with respect to length of the time interval.
Markovian models that encapsulate memory can be constructed, for example, by
introducing a recurrent neural network \cite{levine2022framework}, or by the
use of delay embedding \cite{sauer1991embedology}; such Markovian models are more
computationally expedient. 
Temporally non-local modeling has received significant recent attention~\cite{ma2018model,wang2020recurrent,charalampopoulos2021machine,lin2021data}. If diffusion/advection mechanisms are present in the resolved system, memory effects of any state variable would manifest themselves in the state variables as a spatial non-locality. For this reason non-local models with a flexible enough kernel could potentially be used to capture
memory effects, without significantly increasing the computational costs.

\section{Lorenz 96 Systems as Illustrative Examples}
\label{sec:examples}

For our simulation studies of structural model error, we consider two variants of the celebrated Lorenz 96 model \cite{lorenz1996predictability}, described
in what follows.

\subsection{Lorenz Models}
\label{ssec:LMC}

We consider a multiscale Lorenz 96 model, together with a single-scale 
companion model, to illustrate the principles and algorithms described in the subsequent sections. In each case, we use an untruncated version of the model as the true data-generating model and a truncated version as the model in which structural error models are to be learned.

\subsubsection{Multiscale Lorenz 96 Model}
\label{sssec:mL96}
The Lorenz 96 multi-scale system \cite{lorenz1996predictability} describes the evolution of a simplified atmospheric flow, which is periodic along latitude circles (space). It does so through one set of slow variables, $x_k$ ($k=1,\dots,K$), coupled to a set of fast variables, $z_{j,k}$ ($j=1, \dots, J$), whose indices label space coordinates:
\begin{equation}
\label{eq:ml96}
    \begin{aligned}
    \dot{x}_k&=-x_{k-1}(x_{k-2}-x_{k+1})-x_k+F-hc\bar{z}_k,  \\
    \frac{1}{c}\dot{z}_{j,k}&=-bz_{j+1,k}(z_{j+2,k}-z_{j-1,k})-z_{j,k}+\frac{h}{J}x_k.
    \end{aligned}
\end{equation}
Reflecting the periodicity along latitude circles, the variables are periodic in their indices, with 
\begin{equation}
    x_{k+K}   = x_k, \qquad z_{j,k+K} = z_{j,k}, \qquad z_{j+J,k} = z_{j,k+1}.
\end{equation}
The coupling term $hc\bar{z}_k$ describes the impact of the fast dynamics on the slow dynamics, with only the average
\begin{equation}\label{eq:ml96_avg}
\bar{z}_k=\frac{1}{J}\sum_{j=1}^J z_{j,k}
\end{equation}
of the fast variables affecting the slow variables. To generate data, we work with the parameter choices $K=36$, $J=10$, and $F=b=10$ \cite{lorenz1996predictability,schneider2017earth}. The choices of $h$ and $c$ are summarized in Subsection~\ref{ssec:resultsL} for different cases.

To study how to model structural errors, we consider a coarse-grained system in which we only simulate approximate versions $X_k$ of the slow variables $x_k$, neglecting the fast variables. The approximate slow variables are governed by the system, 
\begin{equation}
\label{eq:ml96c}
\begin{aligned}
\dot{X_k}&=-X_{k-1}(X_{k-2}-X_{k+1})-X_k+F + \delta(X_k, X_k^-;\theta_\struct),\\
X_{k+K}&=X_k,
\end{aligned}
\end{equation}
for $X_k^-=(X_{k-d},\cdots, X_{k-1}, X_{k+1},\cdots, X_{k+d}).$
Here, $\delta(\cdot)$ is the error model that accounts for the missing multiscale interactions.
If there is no dependence on $X_k^-$, the model is local; otherwise, we allow
for non-local dependency with a stencil of width $d$ on either side of $X_k$.
If specified correctly, the model error ensures that the resolved variables $X_k$ of the coarse-grained system \eqref{eq:ml96c} approximate the variables $x_k$ of the full data generating system \eqref{eq:ml96}--\eqref{eq:ml96_avg}. We will use data generated with the full system \eqref{eq:ml96}--\eqref{eq:ml96_avg} to learn about the error model $\delta(\cdot)$ in the coarse-grained system \eqref{eq:ml96c}.

As data $y=\cH[x(\cdot)]$ we consider
\[
\cH[x(\cdot)]=\bbE \cF(x(\cdot)),
\]
where $\bbE$ denotes expectation with respect to the
stationary distribution of $x(t)$ and $\cF:\cX_{\textrm{traj}} \mapsto \bbR^q$ is a function on the
space of solution trajectories (i.e., $\cX_{\textrm{traj}}: 0^+ \mapsto \cX$). In this work, we use moments of the vector $x$ and the averaged auto-correlation function as data and employ a finite-time average to approximate the expectation $\mathbb{E}$:
\begin{enumerate}[(i)]
\item We will use $m^\textrm{th}-$moments of the vector $x$, i.e., $\cF_m(x) = \Pi_{k \in M}x_k$:
\[\cH[x(\cdot)] \approx \frac{1}{T} \int_0^T \Pi_{k \in M}x_k(t) \diff t,\]
where $x_k$ denotes the $k^\textrm{th}$ element of vector $x$, and $M$ is a subset of size $m$ comprising indices (repetition allowed) from
$\{1,\cdots, K\}$. 
\item We will also use autocorrelation function $\cF_{ac}(x(\cdot))=x(t+\tau) \otimes x(t)$:
\[\cH[x(\cdot)] \approx \frac{1}{T} \int_0^T x(t+\tau) \otimes x(t) \diff t.\]
In this work, we only consider the autocorrelation of the same element in the vector $x$, i.e., $x_k(t+\tau)x_k(t)$.
\end{enumerate}

\subsubsection{Single-scale Lorenz 96 Model}
\label{sssec:sL96}

The single-scale Lorenz 96 system does not include the fast variables. We use it to illustrate the combined use of direct and indirect data and the advantage of enforcing conservation constraints in error models. The single-scale Lorenz 96 system describes the evolution of the $x_k$ variables alone,
\begin{equation}
\label{eq:sl96}
    \begin{aligned}
    \dot{x}_k&=-x_{k-1}(x_{k-2}-x_{k+1})-x_k+F, \\
    \quad x_{k+K} &= x_k,
    \end{aligned}
\end{equation}
and we use it to generate data in the setting where $K=36$ and $F=10$. 

As the truncated system, we will only assume that we know the linearized part of the dynamics, resulting in an approximate model of the form
\begin{equation}
\label{eq:sl96c}
\begin{aligned}
\dot{X_k}&= -X_k+F + \delta(X_{k-2},X_{k-1},X_{k+1},X_{k+2};\theta_\struct),\\
X_{k+K}&=X_k.
\end{aligned}
\end{equation}

Here, $\delta$ is the error model that models the missing quadratic terms; we note that we
postulate the need to learn a single universal function $\delta$ to account for model
error in each component of the equation, reflecting an {\em a priori} assumption
about the homogeneity of the structural error with respect to $k.$ Since the error model  $\delta$ accounts for the unknown convection term and thus should not introduce additional energy, the state variable $X_k$ is excluded from the inputs of $\delta$ in the $k$-th equation.
We will use data generated from the untruncated system \eqref{eq:sl96} to learn about the error
$\delta$ in the truncated system \eqref{eq:sl96c}.
As data $y=\cH[x(t)]$, we employ the same types of data (i.e., moments and autocorrelation of the vector $x$) as described in Section~\ref{ssec:sL96}.

\subsection{Numerical Results for Lorenz Models}
\label{ssec:resultsL}

Before presenting detailed numerical results for Lorenz systems, we summarize several highlights of our numerical results.
\begin{enumerate}
    \item For a multiscale system with clear scale separation, local deterministic error models using either direct or indirect data lead to satisfactory model fits. Detailed results are presented in Figs.~\ref{fig:mL96_c_10_direct_training} and~\ref{fig:mL96_c_10_indirect_learning}.
    \item For a multiscale system with less clear scale separation, local deterministic error models using direct data or indirect data do not lead to satisfactory model fits. Detailed results are presented in Figs.~\ref{fig:mL96_c_3_direct_training} and~\ref{fig:mL96_c_3_ODE}. However non-local or stochastic error models do lead to satisfactory fits. Detailed results are presented in Figs.~\ref{fig:mL96_c_3_ODE_nonlocal} to~\ref{fig:mL96_c_3_SDE_multiplicative}.
    \item For the single-scale Lorenz model, we show how an energy constraint can be incorporated into the EKI learning framework and leads to enhanced calibration of the error model. Detailed results are presented in Figs.~\ref{fig:sL96_c_10_direct_training} to~\ref{fig:sL96_c_10_constrained}.
\end{enumerate}

Here, we investigate the long-term behavior of the trained models for the Lorenz systems via the invariant measure. For the numerical examples, the invariant measure corresponds to the probability density function of a system state variable, when simulating the dynamical system for a long enough time so that the probability density function does not vary. For the numerical examples with indirect data, the invariant measure is employed as a qualitative indicator for the long-term performance of the trained models. Considering that the indirect data only contain partial information about the true invariant measure, the quantitative comparison of the ensemble mean for the invariant measures may lead to over-interpretation of the results; thus, we only present the ensemble of invariant measures.

\subsubsection{Lorenz 96 Multi-scale Model}
\label{ssec:mL96_ex}
\begin{figure}[!htbp]
  \centering
  \subfloat[Trained model]{\includegraphics[width=0.44\textwidth]{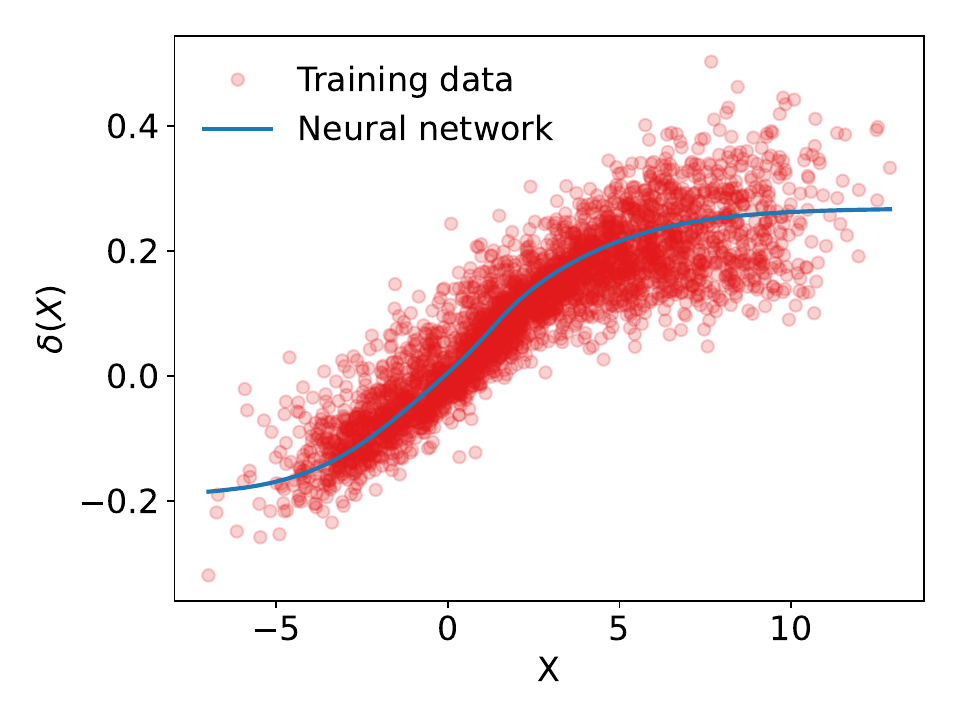}}
  \subfloat[Invariant measure]{\includegraphics[width=0.44\textwidth]{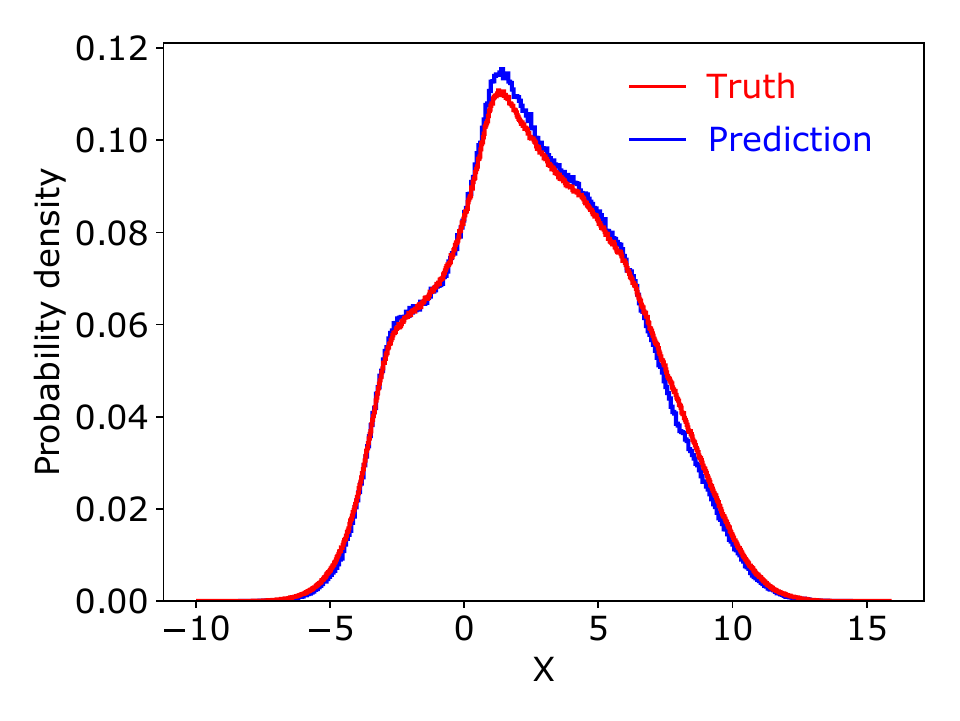}}
    \caption{Direct training of the error model ($c=10$) using a neural network, with results for (a) the trained error model and (b) the invariant measure.}
  \label{fig:mL96_c_10_direct_training}
\end{figure}

\begin{figure}[!htbp]
  \centering
  \subfloat[Data comparison]{\includegraphics[width=0.44\textwidth]{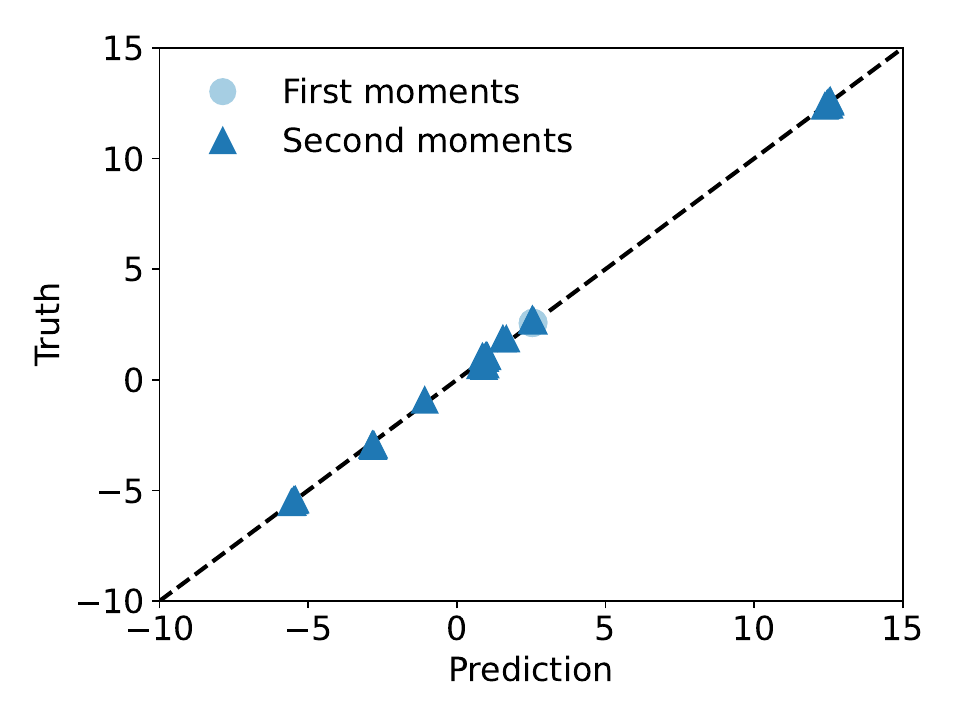}}
  \subfloat[Invariant measure]{\includegraphics[width=0.44\textwidth]{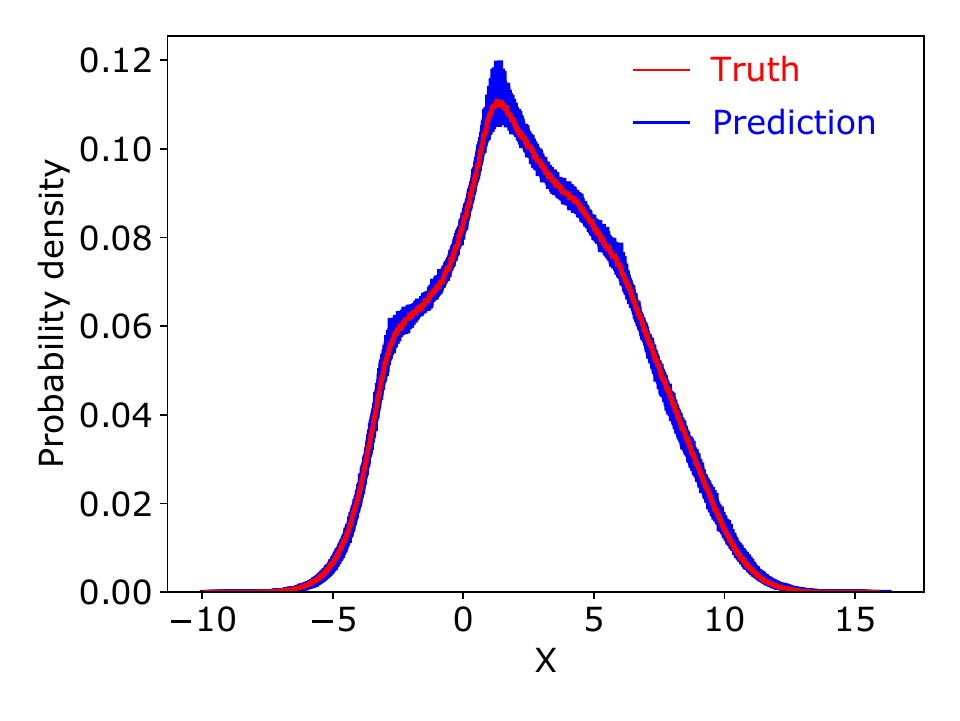}}
    \caption{Indirect training of the error model ($c=10$) using dictionary learning, with the results for (a) first and second moments and (b) the invariant measure. The results with a GP and a neural network have similar performance and are omitted.}
  \label{fig:mL96_c_10_indirect_learning}
\end{figure}

\begin{figure}[!htbp]
  \centering
  \subfloat[Trained model]{\includegraphics[width=0.44\textwidth]{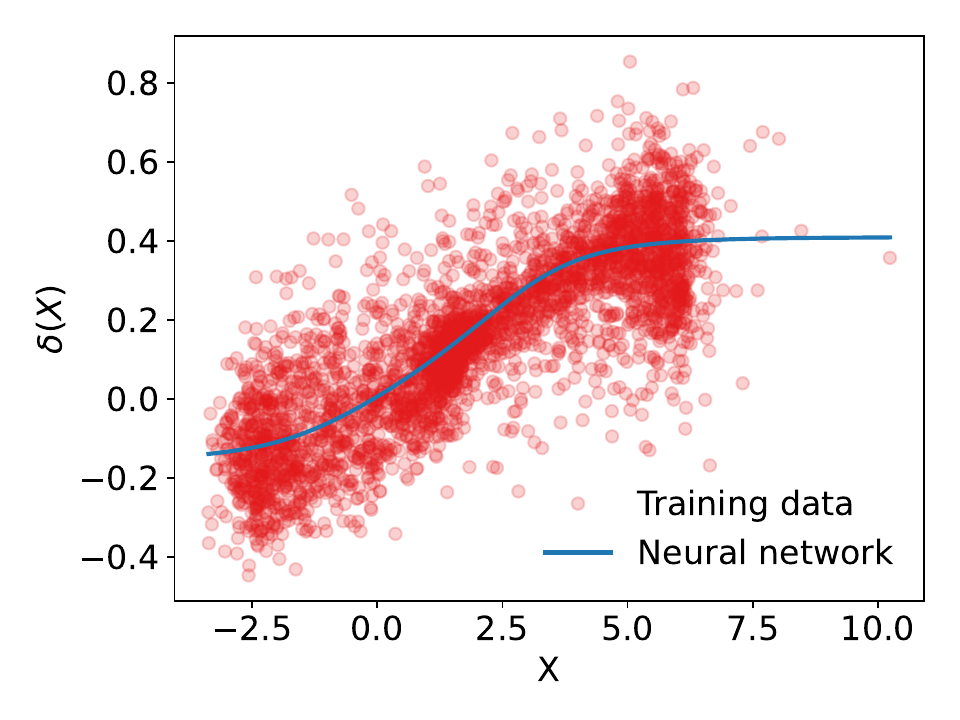}}
  \subfloat[Invariant measure]{\includegraphics[width=0.44\textwidth]{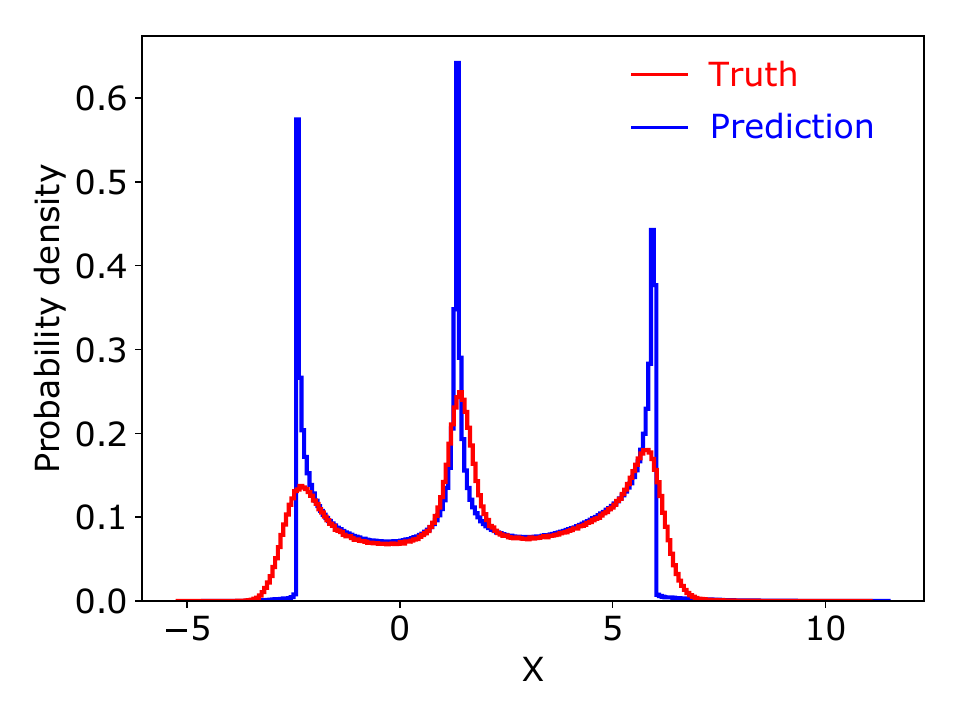}}
    \caption{Direct training of the error model ($c=3$) using a neural network, with results for (a) the trained error model and (b) the invariant measure.}
  \label{fig:mL96_c_3_direct_training}
\end{figure}

\begin{figure}[!htbp]
  \centering
  \subfloat[Data comparison]{\includegraphics[width=0.44\textwidth]{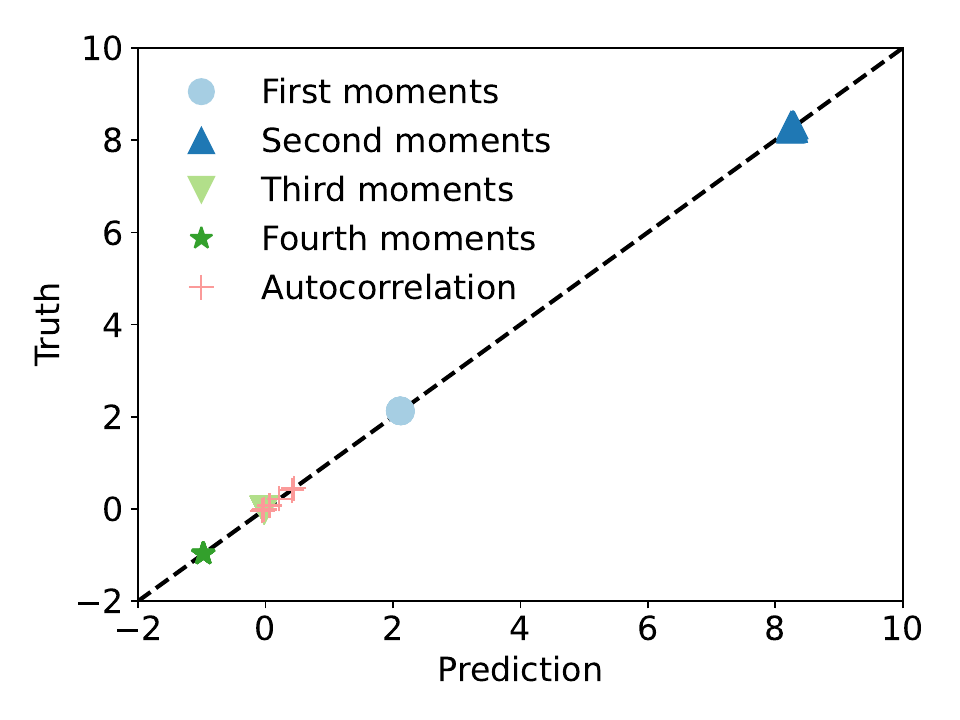}}
  \subfloat[Invariant measure]{\includegraphics[width=0.44\textwidth]{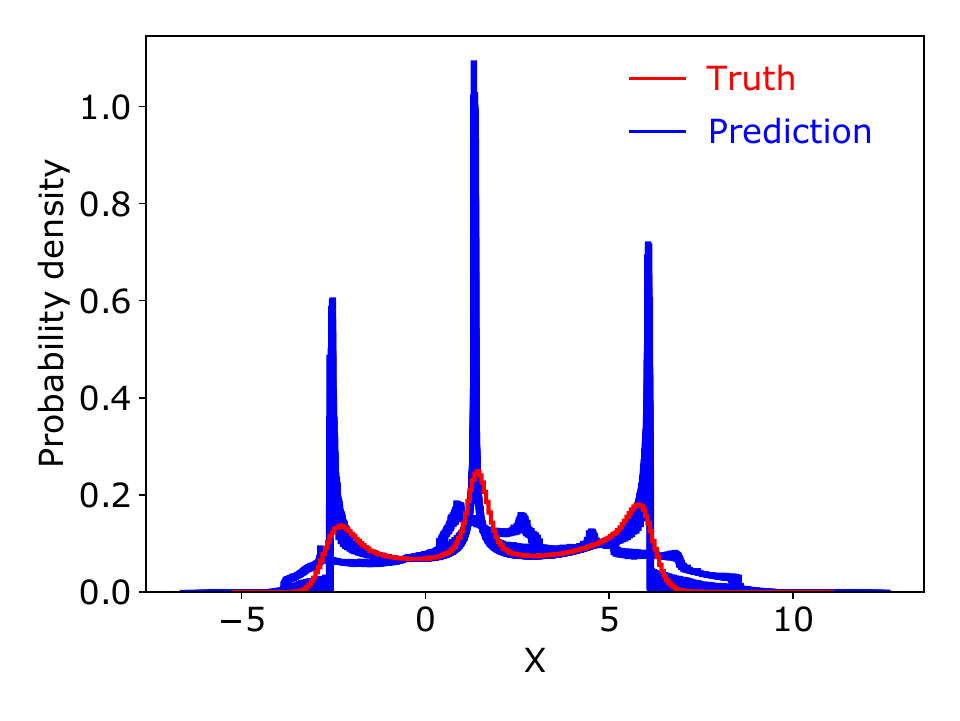}}
    \caption{Indirect training of the error model ($c=3$) using a deterministic model (local), with the results for (a) the first four moments and autocorrelation, and (b) the invariant measure.}
  \label{fig:mL96_c_3_ODE}
\end{figure}

\begin{figure}[!htbp]
  \centering
  \subfloat[Data comparison]{\includegraphics[width=0.44\textwidth]{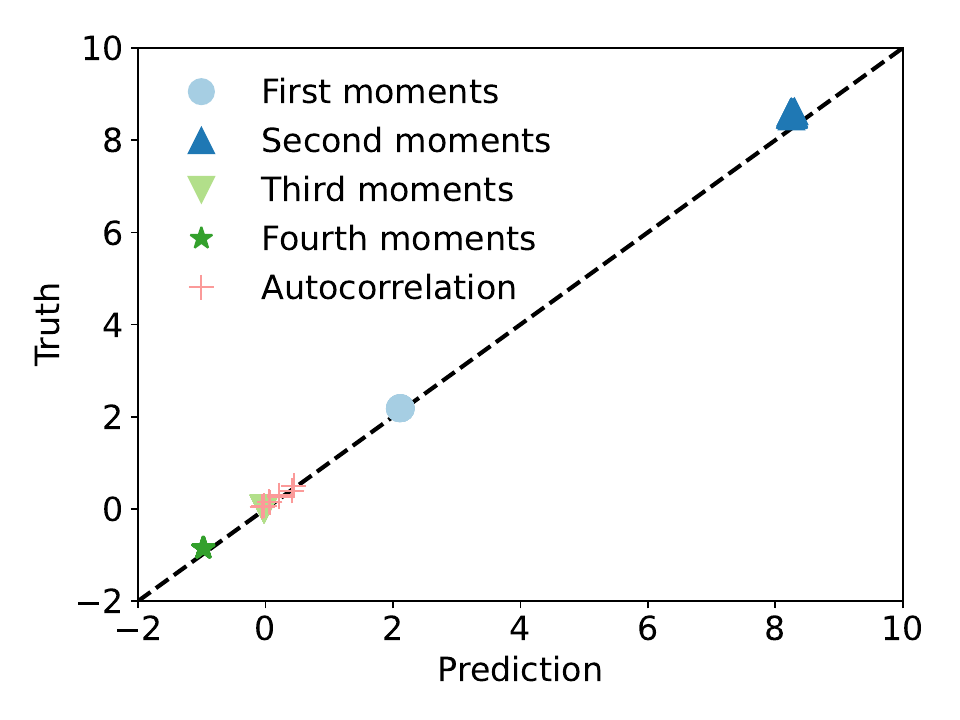}}
  \subfloat[Invariant measure]{\includegraphics[width=0.44\textwidth]{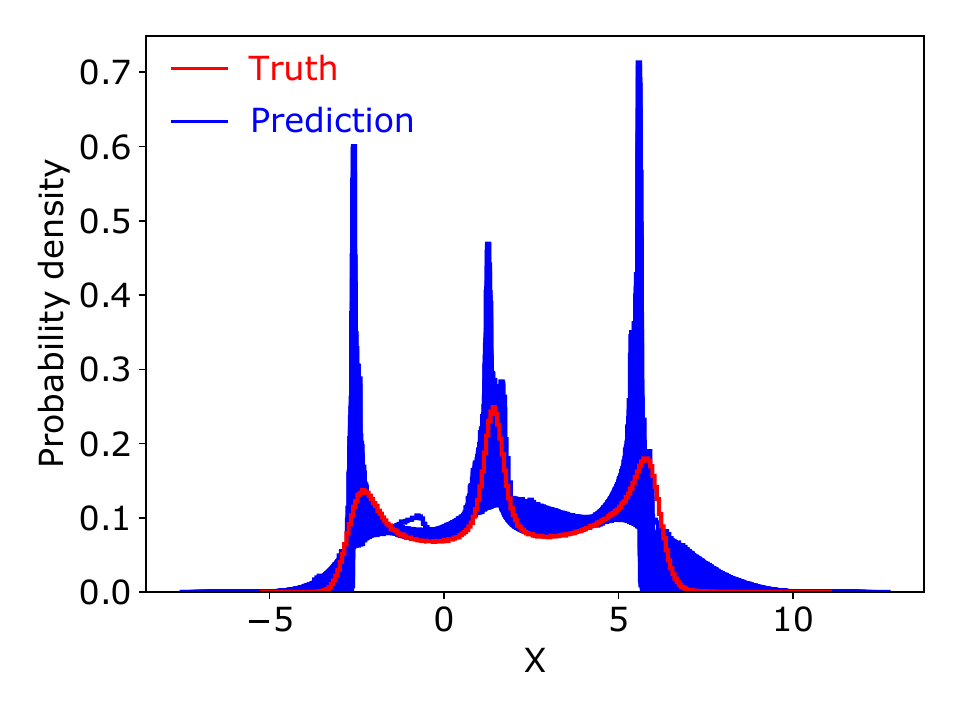}}
    \caption{Indirect training of the error model ($c=3$) using a non-local deterministic model, with results for (a) the first four moments and autocorrelation, and (b) the invariant measure.}
  \label{fig:mL96_c_3_ODE_nonlocal}
\end{figure}

\begin{figure}[!htbp]
  \centering
  \subfloat[Data comparison]{\includegraphics[width=0.44\textwidth]{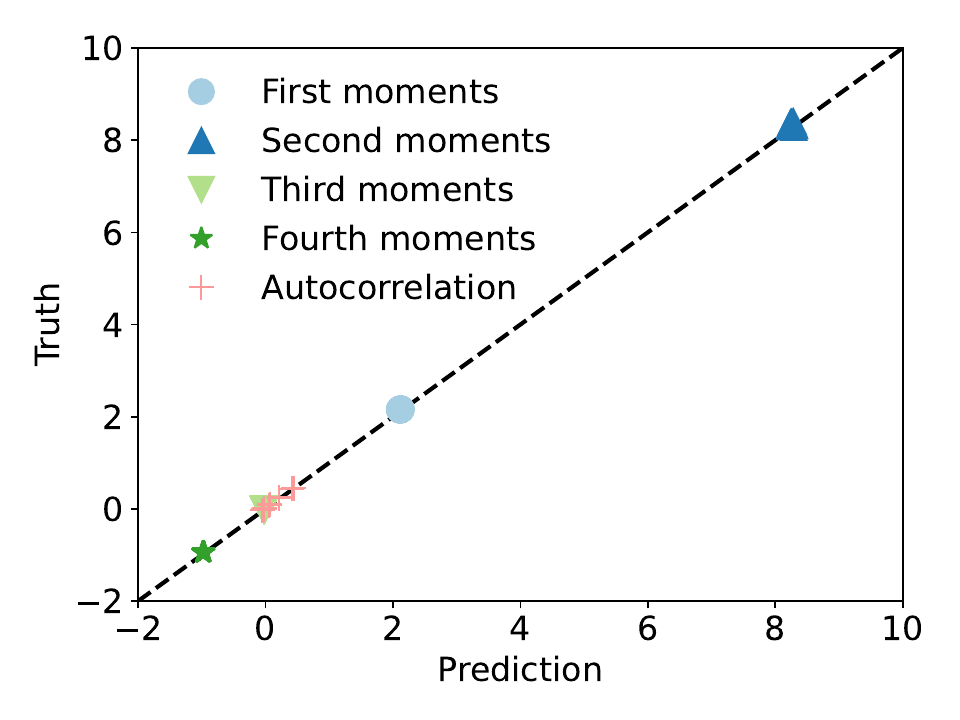}}
  \subfloat[Invariant measure]{\includegraphics[width=0.44\textwidth]{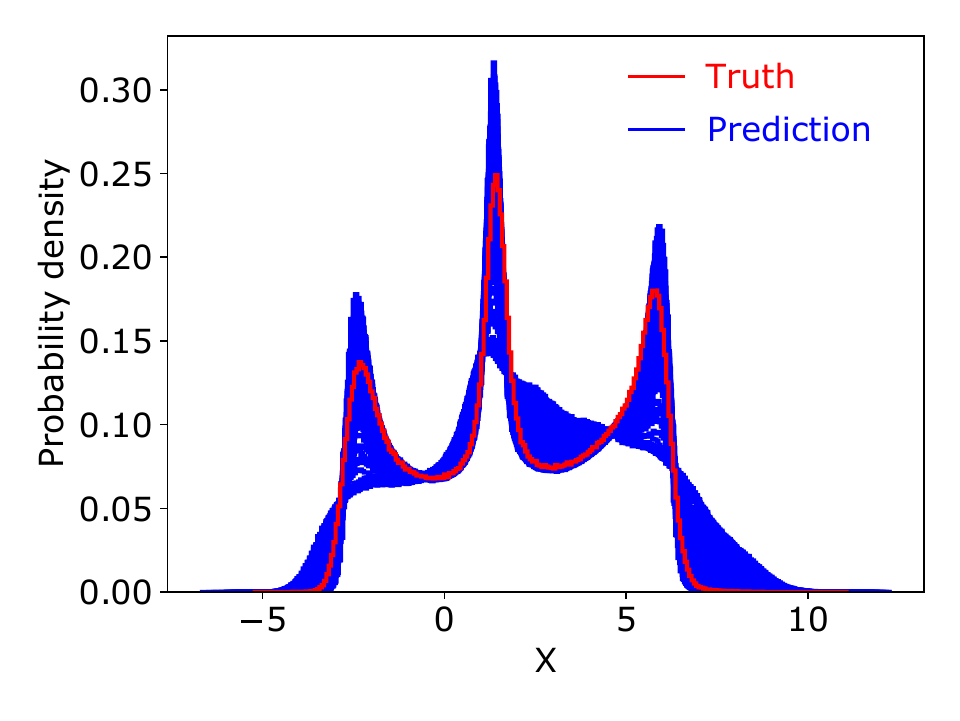}}
    \caption{Indirect training of the error model ($c=3$) using stochastic model with additive noise, with the results for (a) the first four moments and autocorrelation, and (b) the invariant measure.}
  \label{fig:mL96_c_3_SDE_additive}
\end{figure}

\begin{figure}[!htbp]
  \centering
  \subfloat[Data comparison]{\includegraphics[width=0.44\textwidth]{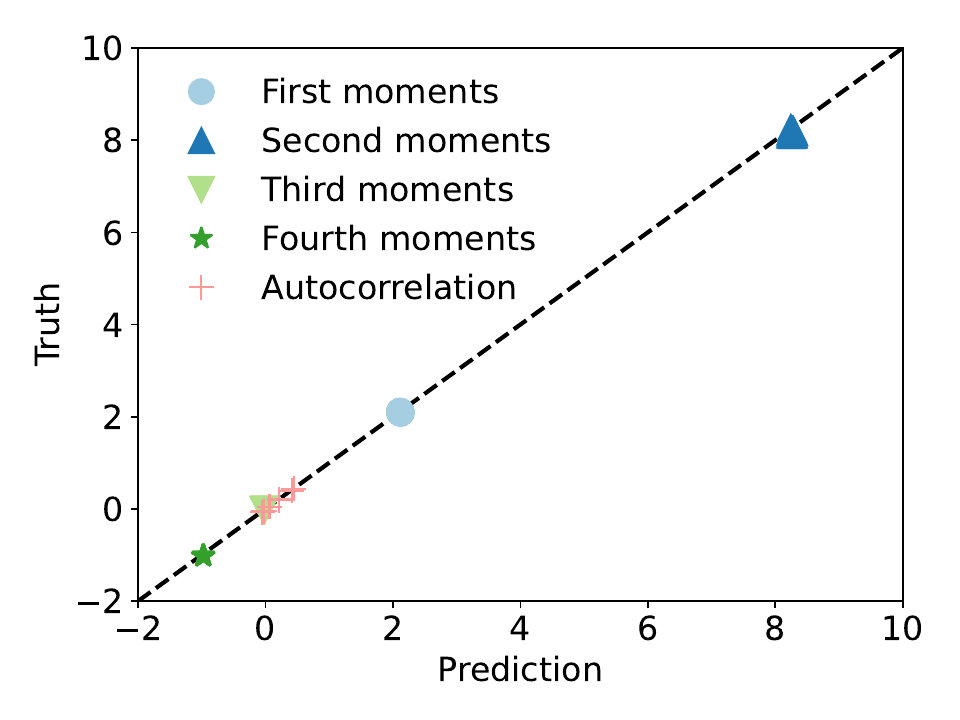}}
  \subfloat[Invariant measure]{\includegraphics[width=0.44\textwidth]{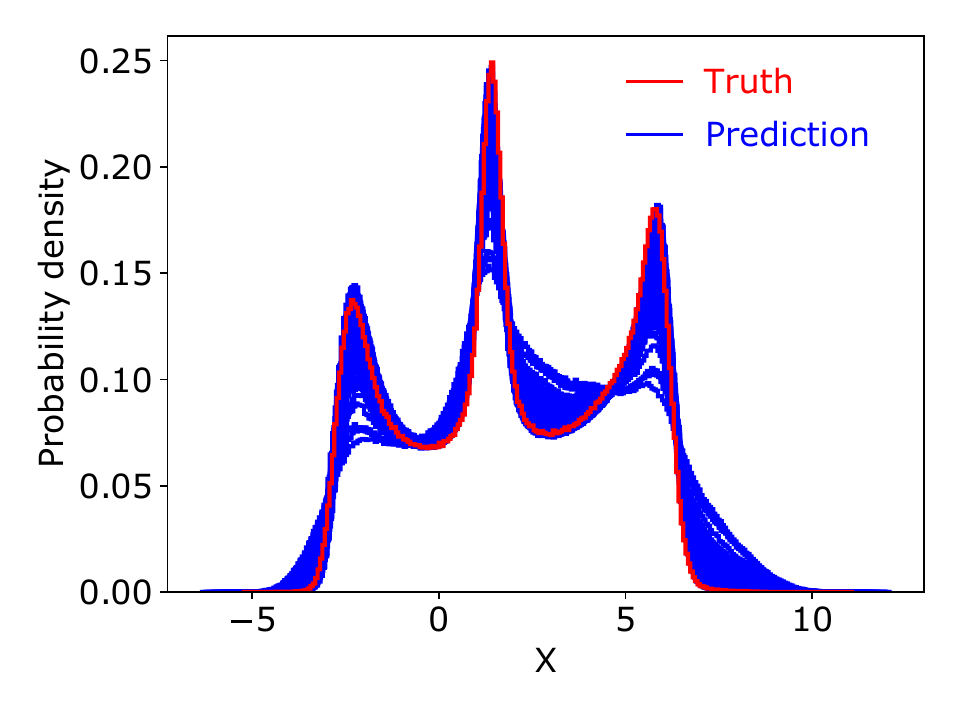}}
    \caption{Indirect training of the error model ($c=3$) using a stochastic model with multiplicative noise, with the results for (a) the first four moments and autocorrelation, and (b) the invariant measure.}
  \label{fig:mL96_c_3_SDE_multiplicative}
\end{figure}

We first study the multi-scale Lorenz 96 system. The numerical examples for the multi-scale Lorenz 96 system are as follows:
\begin{enumerate}[(i)]
    \item For $c=10$ and $h=1$ in the multi-scale Lorenz 96 system, a fully-connected neural network is trained as a local deterministic error model $\delta(X_k)$ using direct data ($\{x_k,hc\overline{z}_k\}$), and a dictionary-learning-based model is trained using indirect data (first and second moments of the slow variable $x_k$). The direct data comprise 36000 data points. For the indirect data in this example, we assume partial observations of the first eight slow variables and include cross-terms of second moments (i.e., $\mathbb{E}(x_ix_j)$ for different $i$ and $j$). The indirect data comprise 44 data points. The results are presented in Figs.~\ref{fig:mL96_c_10_direct_training} and~\ref{fig:mL96_c_10_indirect_learning}. 
    \item For $c=3$ and $h=10/3$ in the multi-scale Lorenz 96 system, the scale separation between fast and slow variables becomes smaller and thus leads to a more challenging case. In this case, a fully-connected neural network is trained as the local deterministic error model $\delta(X_k)$ using either direct data ($\{x_k,hc\overline{z}_k\}$) or indirect data (first to fourth moments of the slow variable $x_k$ and the autocorrelation of $x_k$). For the indirect data in this and the next examples, we enable the full observation of all 36 slow variables and preclude the use of all cross-terms of second to fourth moments. The direct data comprise 36000 data points, and the indirect data 154 data points. The results are presented in Figs.~\ref{fig:mL96_c_3_direct_training} and~\ref{fig:mL96_c_3_ODE}.
    \item For $c=3$ and $h=10/3$ in the multi-scale Lorenz 96 system, we train a non-local deterministic error model $\delta(X)=\sum_{k^\prime} \delta(X_{k^\prime}) \mathcal{C}(k-k^\prime;\theta_\nonlocal)$, a local stochastic error model with additive noise $\delta(X_k) + \sqrt{\sigma^2}\dot{W_k}$, and a local stochastic error model with multiplicative noise $\delta(X_k) + \sqrt{\sigma^2(X_k)}\dot{W_k}$, using indirect data (first to fourth moments of the slow variable $x_k$ and the autocorrelation of $x_k$). The results are presented in Figs.~\ref{fig:mL96_c_3_ODE_nonlocal} to~\ref{fig:mL96_c_3_SDE_multiplicative}.
\end{enumerate}

Figure~\ref{fig:mL96_c_10_direct_training}a presents the direct data $\{x_k,hc\bar{z}_k\}$. Based on direct data, a regression model $\delta(X_k)$ can be trained and then used to simulate the dynamical system of $X_k$ in \eqref{eq:ml96c}. In this work, we train such a regression model using a fully-connected neural network with two hidden layers (five neurons at the first hidden layer and one neuron at the second hidden layer). It can be seen in Fig.~\ref{fig:mL96_c_10_direct_training}a that the trained model captures the general pattern of the training data. We also simulate the dynamical system in \eqref{eq:ml96c} for a long time trajectory and compare the invariant measure of $X_k$ with the true system in \eqref{eq:ml96}. As shown in Fig.~\ref{fig:mL96_c_10_direct_training}b, we obtain a good agreement between the invariant measures of the modeled system and the true system.

Because direct data $\{x_k,hc\bar{z}_k\}$ may not be accessible in some applications, we also explore the use of indirect data to calibrate the error model $\delta(X_k)$. In this example, the first and second moments of the first eight components of $x_k$ are used for the calibration. We tested different approaches that parameterize the error model $\delta(X_k)$, including dictionary learning (Fig.~\ref{fig:mL96_c_10_indirect_learning}), GPs and neural networks. The error model based on dictionary learning has the form $\delta(X_k) = \sum_{i=1}^2 \alpha_{i} \phi_i(X_k)$, where we choose the basis function dictionary $\phi_i(X_k) \in \{\tanh(\beta_1 X_k), \tanh(\beta_2 X_k^2)\}$. Therefore, we have $\{\alpha_i,\beta_i\}_{i=1}^2$ as unknown parameters to be learned. Instead of using polynomial basis functions, we have introduced the hyperbolic tangent function $\tanh(\cdot)$ to enhance the numerical stability. The error model based on a GP has the form $\delta(X_k) = \sum_{j=1}^7 \alpha_j \mathcal{K}(X_k^{(j)}, X_k;\ \psi)$, where we chose the $X_k^{(j)}$ as seven fixed points uniformly distributed in $[-15,15]$, and $\mathcal{K}$ as a squared exponential kernel with unknown constant hyper-parameters $\psi=\{\sigma_\mathrm{GP},\ell\}$, where $\sigma_\mathrm{GP}$ denotes the standard deviation and $\ell$ the length scale of the kernel. The results with a GP and with a neural network are similar to the ones with dictionary learning in Fig.~\ref{fig:mL96_c_10_indirect_learning} and are omitted here. The calibrated models in all three tests lead to good agreement in both data and invariant measure, and the performance of the calibrated model is not sensitive to the specific choice of parameterization approaches.

Although the performance of the calibrated error model is not sensitive to either the types of data or the parameterization approaches for this numerical example, the specific choices made in constructing and calibrating error models are still important, and even more so for more challenging scenarios, e.g., when the resolved and unresolved degrees of freedom have less scale separation. 

To illustrate the advantage of using indirect data and stochastic/non-local error models, we study a more challenging scenario where the scale separation between $x_k$ and $y_{j,k}$ in \eqref{eq:ml96} is narrower, by setting $h=10/3$ and $c=3$. It can be seen in Fig.~\ref{fig:mL96_c_3_direct_training}a that the general pattern of the direct data is still captured by the trained error model $\delta(X_k)$. However, the comparison of the invariant measures in Fig.~\ref{fig:mL96_c_3_direct_training}b shows that the long-time behaviour of the trained model does not agree with the true system, indicating the limitation of using only direct data for the calibration.

We further investigate the use of indirect data. Specifically, the first four moments of $X_k$ and ten points sampled from the averaged autocorrelation function of $X_k$ are used as training data. Figure~\ref{fig:mL96_c_3_ODE} presents the results for the calibrated local model $\delta(X_k)$. It can be seen in Fig.~\ref{fig:mL96_c_3_ODE}a that the trained error model agrees with the training data, while the invariant measures in Fig.~\ref{fig:mL96_c_3_ODE}b still differ between the calibrated and the true systems, indicating overfitting of the training data.

To avoid the overfitting in Fig.~\ref{fig:mL96_c_3_ODE}, we calibrate a non-local error model, as discussed in Section \ref{ssec:nonlocal}. Compared to the results for the local error model, it can be seen in Fig.~\ref{fig:mL96_c_3_ODE_nonlocal} that the invariant measure of the calibrated system agrees better with the true system, which indicates that the closure model $\delta(\cdot)$ in \eqref{eq:ml96c} with non-local effects represents a better closure for unresolved scales if there is a less clear scale separation between resolved and unresolved scales. The non-local model tends to be more flexible than the local model. Figure~\ref{fig:mL96_c_3_ODE}a demonstrates that the local model already achieves a good agreement in data space with the true system. It is likely that more than one possible non-local model can achieve a comparable performance of matching the indirect data. However, simulating those different non-local models for a much longer time can lead to more noticeable differences in their invariant measures, which provides a possible explanation for the larger variability within the ensemble in Fig.~\ref{fig:mL96_c_3_ODE_nonlocal}b. It should be noted that the key challenge of this example is the relatively large variability of the true closure term in Fig.~\ref{fig:mL96_c_3_direct_training}a. Without accounting for such variability, the local and deterministic model does not correctly capture the invariant measure even in the standard regression setting with an abundance of training data as presented in Fig.~\ref{fig:mL96_c_3_direct_training}b. In the non-local model, two local states with similar values may have noticeably different values of their neighbor states, and thus the non-local model with the local state and its neighbor states as the inputs can address some of the variability in Fig.~\ref{fig:mL96_c_3_direct_training}a, which explains its improved performance relative to to the local models. Considering that there are three peaks in the invariant measure of the true system, the non-local model has a greater chance of having some inputs from the left or right peak than merely having inputs from the central peak. Therefore, the trained model can be more in favor of the left or right peak of the invariant measure, which provides a possible explanation for the pattern of peaks presented in Fig.~\ref{fig:mL96_c_3_ODE_nonlocal}b.

We also explored learning a stochastic error model for this example. The neural networks used in the stochastic error models have 5 neurons in the first hidden layer and one neuron in the second layer. The hidden layer neurons have a sigmoid activation function, and the output layer neurons have no nonlinear activation function. For the stochastic error model with an additive noise term, 19 parameters need to be estimated. For the stochastic error model with a multiplicative noise term, 36 parameters need to be estimated. Figure~\ref{fig:mL96_c_3_SDE_additive} presents the results of the calibrated system with an additive stochastic error model. Compared to the results for deterministic error models in Figs.~\ref{fig:mL96_c_3_ODE} and~\ref{fig:mL96_c_3_ODE_nonlocal}, we can see that the invariant measure of the calibrated system agrees better with the true system in Fig.~\ref{fig:mL96_c_3_SDE_additive}b. We further test the stochastic error model by also learning a state-dependent diffusion coefficient. As shown in Fig.~\ref{fig:mL96_c_3_SDE_multiplicative}b, the calibrated system achieves better agreement with the invariant measure of the true system, which confirms that increased flexibility in the stochastic error model can help achieve improved predictive performance via training against indirect data. It should be noted that Figs.~\ref{fig:mL96_c_3_ODE_nonlocal} to~\ref{fig:mL96_c_3_SDE_multiplicative} do not display a single converged invariant measure. This arises because
the calibrated parameters vary across the ensemble; as a result, we obtain a family of structural error models that all fit the data with similar accuracy. We surmise that this is caused by the fact that the indirect data contain only partial information about the invariant measure. Nonetheless, the fits are all far superior to those obtained with the local deterministic model. 

\subsubsection{Lorenz 96 Single-scale Model}
\label{ssec:sL96}
\begin{figure}[!htbp]
  \centering
  \includegraphics[width=0.44\textwidth]{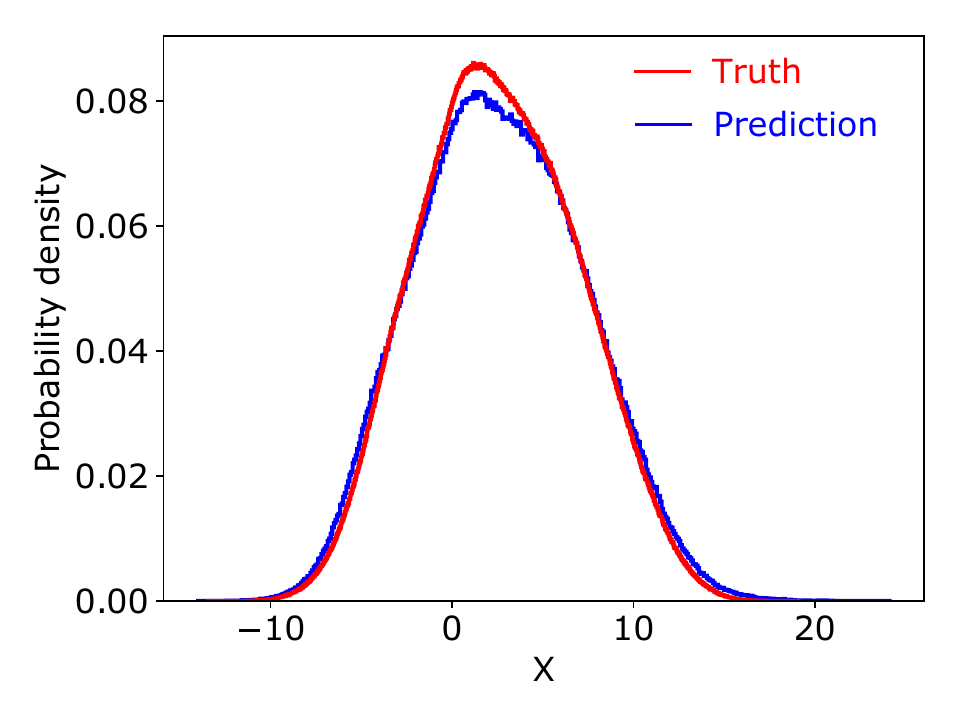}
    \caption{Invariant measures via direct training of the error model ($c=10$) for the single-scale Lorenz 96 system.}
  \label{fig:sL96_c_10_direct_training}
\end{figure}

\begin{figure}[!htbp]
  \centering
  \subfloat[Data comparison]{\includegraphics[width=0.44\textwidth]{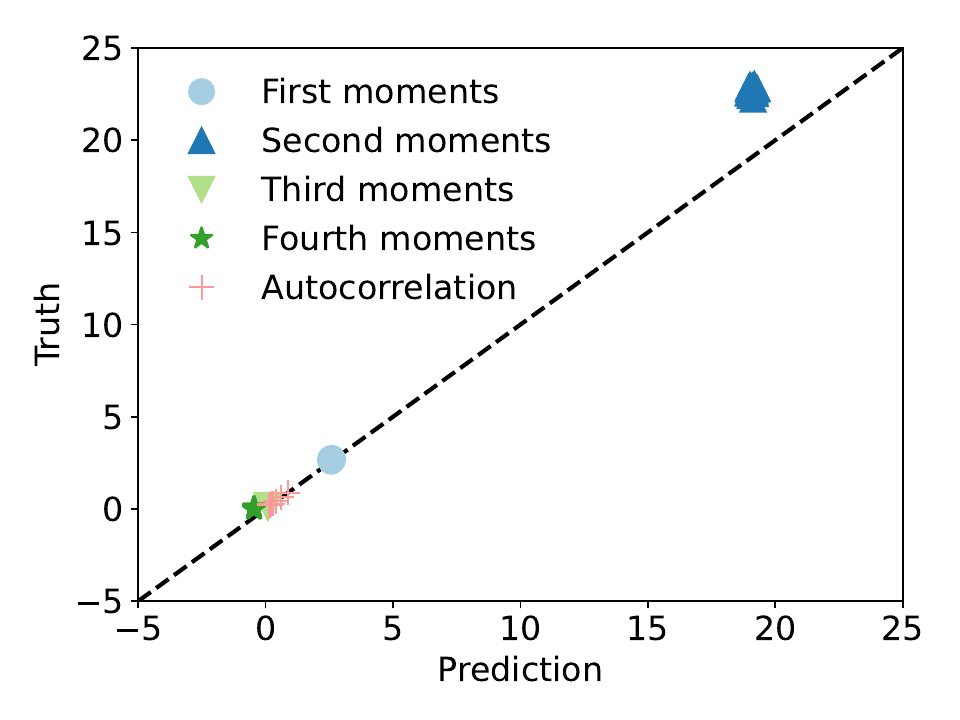}}
  \subfloat[Invariant measure]{\includegraphics[width=0.44\textwidth]{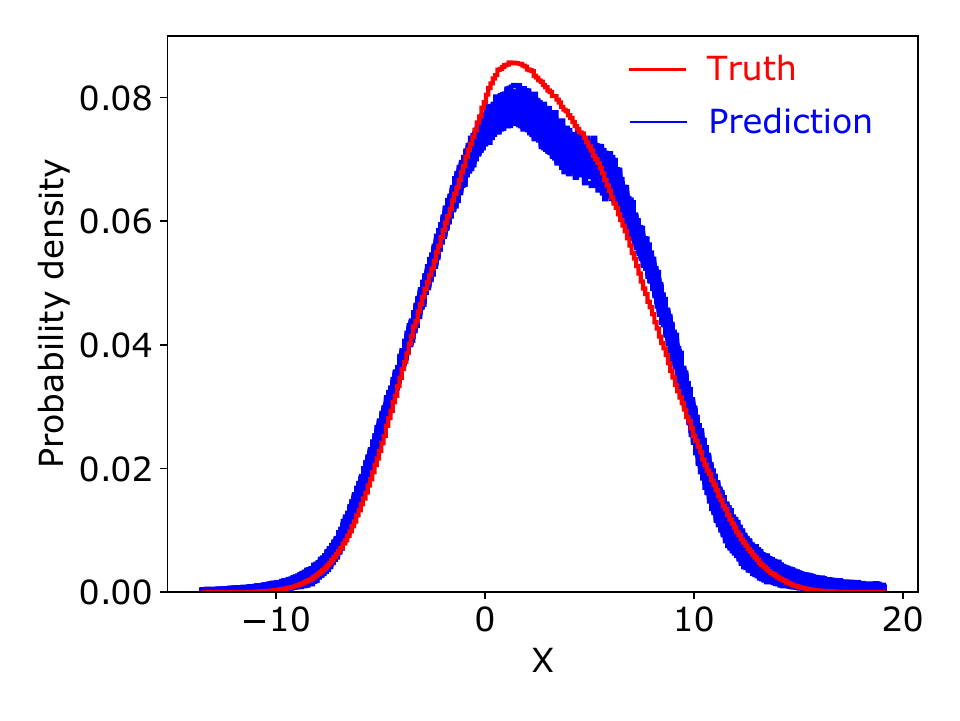}}
    \caption{Results from trained error model ($c=10$) for the single-scale Lorenz 96 system \textit{without} energy conservation constraint, including (a) first four moments and autocorrelation, and (b) invariant measures.}
  \label{fig:sL96_c_10}
\end{figure}

\begin{figure}[!htbp]
  \centering
  \subfloat[Data comparison]{\includegraphics[width=0.44\textwidth]{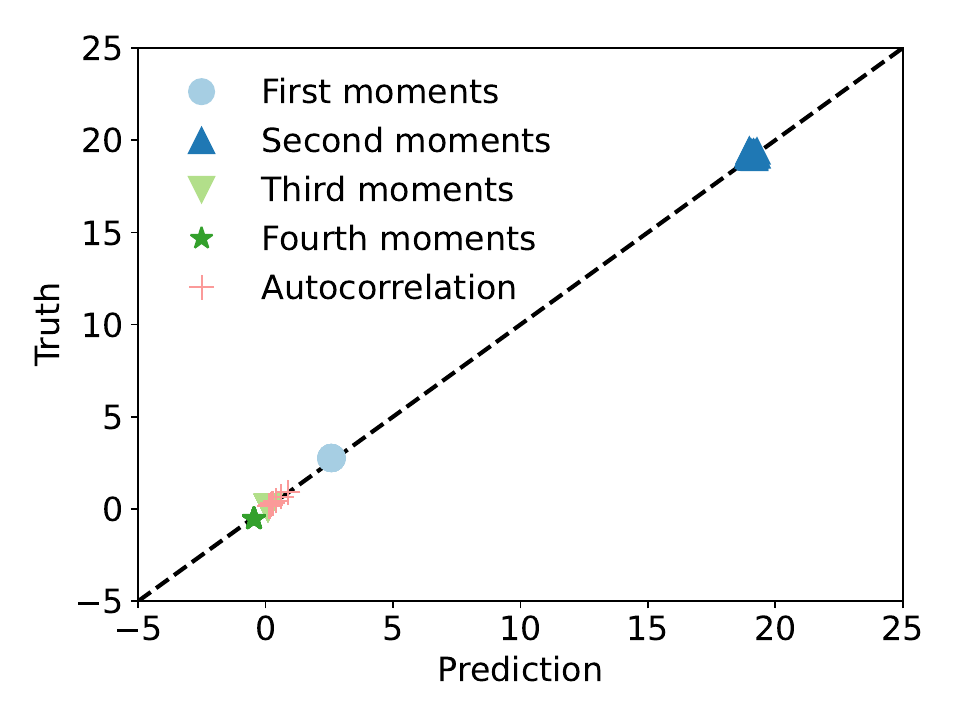}}
  \subfloat[Invariant measure]{\includegraphics[width=0.44\textwidth]{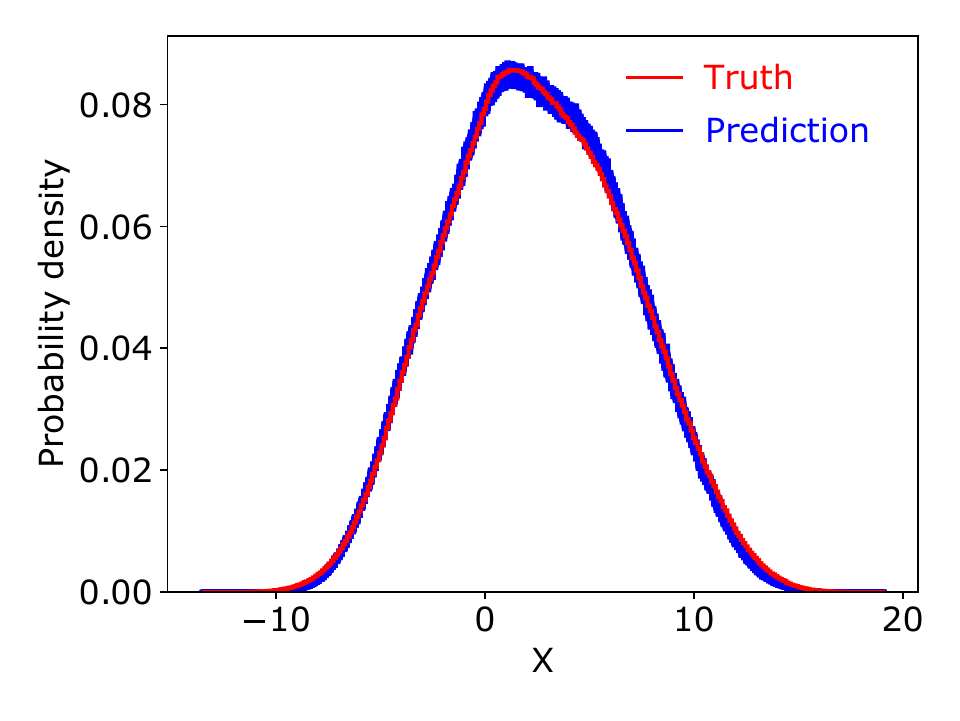}}
    \caption{Results from trained error model ($c=10$) for the single-scale Lorenz 96 system \textit{with} energy conservation constraint, including (a) first four moments and autocorrelation, and (b) invariant measures.}
  \label{fig:sL96_c_10_constrained}
\end{figure}

We studied the Lorenz 96 single-scale system to learn the quadratic term as an error model. Using this numerical example, we demonstrate the merit of combined use of direct and indirect data and the importance of enforcing physical constraints. More specifically, we assume that the quadratic term of the true system in \eqref{eq:sl96} is unknown and then calibrate an error model $\delta(X_{k-2},X_{k-1},X_{k+1},X_{k+2})$ as in \eqref{eq:sl96c}. The size of the direct data is 36000, and the size of the indirect data is 154. The numerical examples for the single-scale Lorenz 96 system are as follows:
\begin{enumerate}[(i)]
    \item We train a fully-connected neural network as an error model $\delta(\cdot)$ using time-series of $x_k$ and the true quadratic term as direct data. The results are presented in Fig.~\ref{fig:sL96_c_10_direct_training}.
    \item We train a fully-connected neural network as an error model $\delta(\cdot)$ using indirect data (first to fourth moments of the state variable $x_k$ and the autocorrelation of $x_k$). The results are presented in Fig.~\ref{fig:sL96_c_10}.
    \item We train a fully-connected neural network as an error model $\delta(\cdot)$ using indirect data (first to fourth moments of the state variable $x_k$ and the autocorrelation of $x_k$) and the energy conservation constraint in~\eqref{eq:constraint_2}. The results are presented in Fig.~\ref{fig:sL96_c_10_constrained}.
\end{enumerate}

Figure~\ref{fig:sL96_c_10_direct_training} presents the comparison of invariant measures between the calibrated system and the true system when a fully connected neural network is used as an error model and is learned from direct data. As seen in Fig.~\ref{fig:sL96_c_10_direct_training}, the calibrated system using direct data still differs noticeably in the invariant measure, indicating a difference from the long-time behavior of the true system.

In order to improve the results of Fig.~\ref{fig:sL96_c_10_direct_training}, we incorporate indirect data about the $x_k$. Specifically, we employ the trained model using direct data as the prior mean of EKI, and we set the prior standard deviation as $30\%$ of the mean values for each unknown coefficient of the error model. We then use EKI to calibrate the error model based on the first four moments of $X_k$ and the ten sampled points from the autocorrelation function of $X_k$. Without enforcing energy conservation of the error model, we can see in Fig.~\ref{fig:sL96_c_10}b that the performance of the calibrated model is similar to the calibrated system using direct data in Fig.~\ref{fig:sL96_c_10_direct_training}. On the other hand, we also performed the calibration based on indirect data and enforced the energy conservation of the error model as discussed in Section~\ref{ssec:constraints}. Considering that the true system conserves quadratic energy at every time, the energy constraint with the form in Eq.~\eqref{eq:constraint_2} is used. As shown in Fig.~\ref{fig:sL96_c_10_constrained}, the calibrated error model with energy conservation leads to a modeled system that better fits the training data and achieves a good agreement with the invariant measure of the true system.

\section{Human Glucose-Insulin Model as Illustrative Example}
\label{sec:examplesGI}

The examples based on Lorenz systems in Section~\ref{sec:examples} illustrate the construction and calibration of internal error models for chaotic and ergodic systems, which are representative of a wide range of applications, e.g., in weather forecasting and climate change prediction, design of hypersonic vehicles, or the control of unmanned vehicles in a turbulent environment. However, there is another large class of applications for which the systems are non-ergodic and the goal is to accurately predict the time series of the system evolution. To achieve this goal for non-ergodic systems, we can no longer neglect the impact of the initial condition. We use a human glucose-insulin model as an example of a non-ergodic system and illustrate how the internal error model can be calibrated based on time series data, using EKI combined with data assimilation techniques.

\subsection{Ultradian Model}
\label{ssec:UM}

We consider the ultradian model of the glucose-insulin system
proposed in \cite{sturisComputerModelMechanisms1991a}. Its primary state variables are the plasma
glucose concentration, $G$, the plasma insulin concentration, $I_{p}$, and the interstitial
insulin concentration, $I_{i}$. These three state variables are augmented
with a three stage delay $(h_1,h_2,h_3)$ which encodes a non-linear delayed hepatic glucose response to plasma insulin levels. The resulting ordinary differential equations have the form:
\begin{subequations}
	\label{eq:ultradian}
	\begin{align}
	   \frac{dI_p}{dt}&=f_1(G)-E\Big(\frac{I_p}{V_p}-\frac{I_i}{V_i}\Big)-\frac{I_p}{t_p} \label{eq:UM1}\\
	   \frac{dI_i}{dt}&=E\Big(\frac{I_p}{V_p}-\frac{I_i}{V_i}\Big)-\frac{I_i}{t_i} \label{eq:UM2}\\
	   \frac{dG}{dt}&=-f_2(G)-f_3(I_i)G+f_4(h_3) +m_G(t) \label{eq:UM3} \\ 
	   \frac{dh_1}{dt}&=\frac{1}{t_d}(I_p-h_1)\label{eq:UM4}\\
	   \frac{dh_2}{dt}&=\frac{1}{t_d}(h_1-h_2)\label{eq:UM5}\\
	   \frac{dh_3}{dt}&=\frac{1}{t_d}(h_2-h_3)\label{eq:UM6}
	\end{align}
\end{subequations}
Here
$m_G(t)$ represents a known rate of ingested carbohydrates appearing in the plasma,
$f_1(G)$ represents the rate of glucose-dependent insulin production,
$f_2(G)$ represents insulin-independent glucose utilization, 
$f_3(I_i)G$ represents insulin-dependent glucose utilization,
and
$f_4(h_3)$ represents delayed insulin-dependent hepatic glucose production.
The functional forms of these parameterized processes are shown in \ref{sec:appendix_ult}.

We represent a potential structural error in this model by removing the final linear term in \eqref{eq:UM1} (i.e., $t_p = \infty$).
We then aim to identify a neural-network-based additive correction term for \eqref{eq:UM1}, denoted by $\delta(I_p, I_i, G, h_1, h_2, h_3; \theta)$.
We calibrate $\delta(\cdot \ ; \ \theta)$ using partial, noisy observations of the physiologic state at times $\{t_k\}_{k=0}^K$.
Specifically, we observe only the glucose state, $G(t)$, (it is the only state variable that is reliably measurable in patients) and add i.i.d.\ unit-Gaussian noise. Thus the observation operator $\cH$ in Eq.~\eqref{eq:ydag} is defined through observation matrix $H \in \bbR^{1 \times 6}$. To be specific $H X(t) = X_3(t)$ where $X(t)=[I_p(t), I_i(t), G(t), h_1(t), h_2(t), h_3(t)]^T$ and then $\cH[X]=\{H X(t_k)\}_{k=0}^K.$ It should be noted that the observation matrix $H$ projects between the full system state space and the observable space and may be used in the definition of a gain matrix $K$ for Kalman 
methods and generalizations such as 3DVAR (See Table 9.1 in \cite{sanz2023inverse}).

We refer to the observed data as
$y=\big[y(t_1), \cdots, y(t_{K})\big]$ and to the solution of the entire
six dimensional dynamical system at the observation times as $\big[x(t_1), \cdots, x(t_{K})\big].$
We infer parameters $\theta$ by minimizing a variant of the loss defined in Eq.~\eqref{eq:ic_DA}
 in which the forward model itself depends on $y$:
\begin{equation}
\label{eq:EKI_3dvar}
\begin{aligned}
    \cL(\theta) = \frac12 \bigl|y-\cG\bigl(\theta; \ y\bigr)\bigr|^2_{\Sigma}.
\end{aligned}
\end{equation}
Specifically $\cG_{k}(\theta; \ y)$ uses sequential data assimilation to combine previous observations $\big[y(t_1), \cdots, y(t_{k-1})\big]$ with parameters $\theta$ to produce a filtered estimate for $x(t_{k-1})$ and, from this, predict the observation $y(t_k)$; see section \ref{ssec:ics}.
The minimization of Eq.~\eqref{eq:EKI_3dvar} is performed using EKI, where each evaluation of $\cG(\theta;y)$ involves running a data assimilation sub-routine (3DVAR \cite{sanz2023inverse} with constant gain $K = [0,0,1,0,0,0]$).

\subsection{Numerical Results for Ultradian Model}
\label{ssec:resultsU}

\begin{figure}[!htbp]
  \centering
  \subfloat[Trained model: $\delta$ vs $I_p$]{\includegraphics[height=0.225\textwidth]{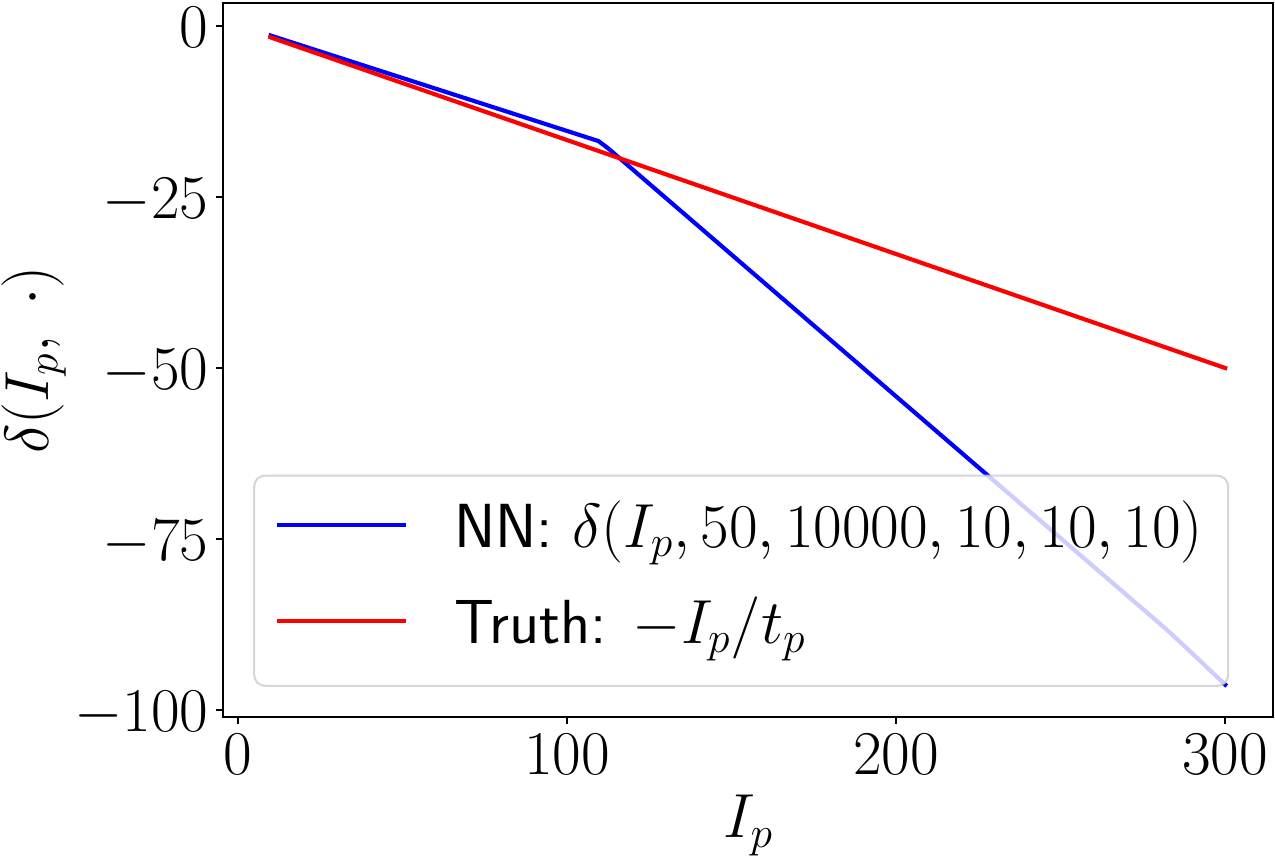} \label{fig:glucoseIp}} 
  \subfloat[Trained model: $\delta$ vs $G$]{\includegraphics[height=0.225\textwidth]{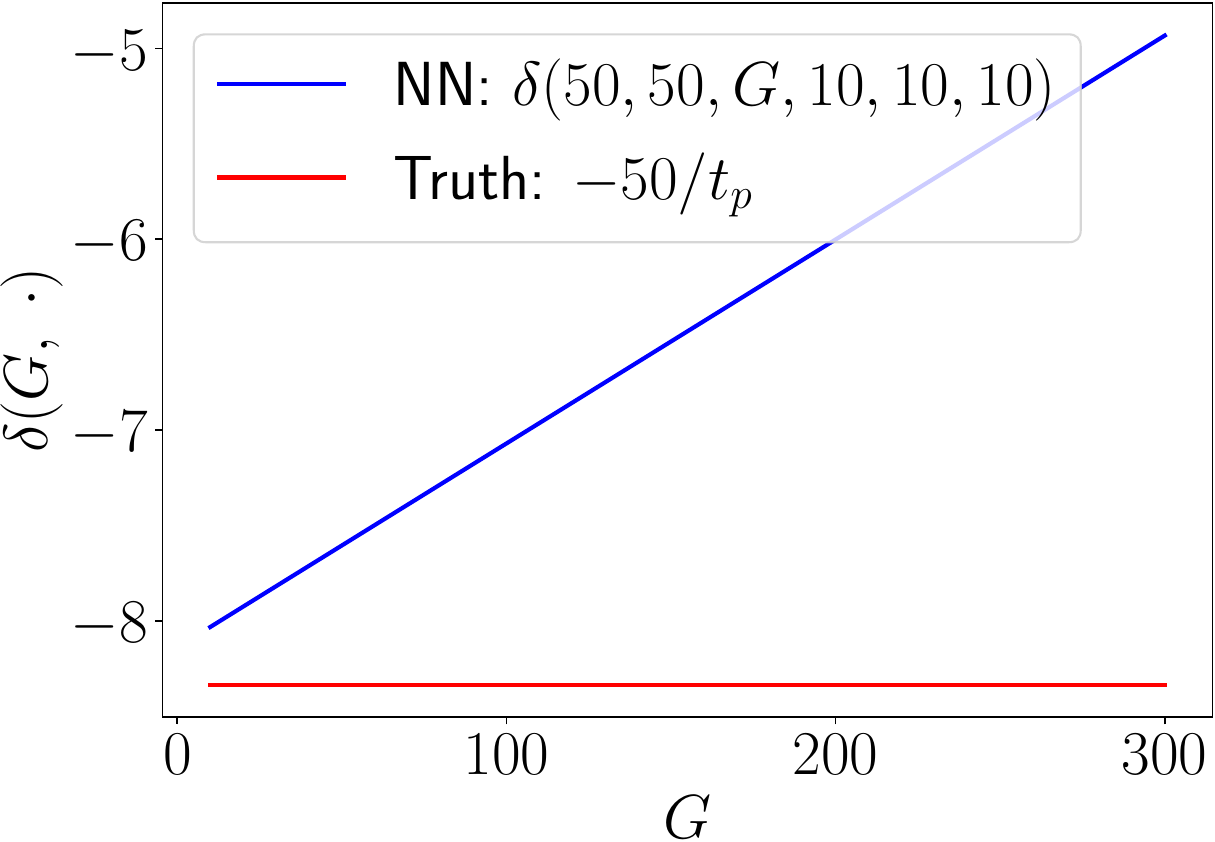} \label{fig:glucoseG}}
  \subfloat[Glucose trajectory fits: $G(t)$]{\includegraphics[height=0.225\textwidth]{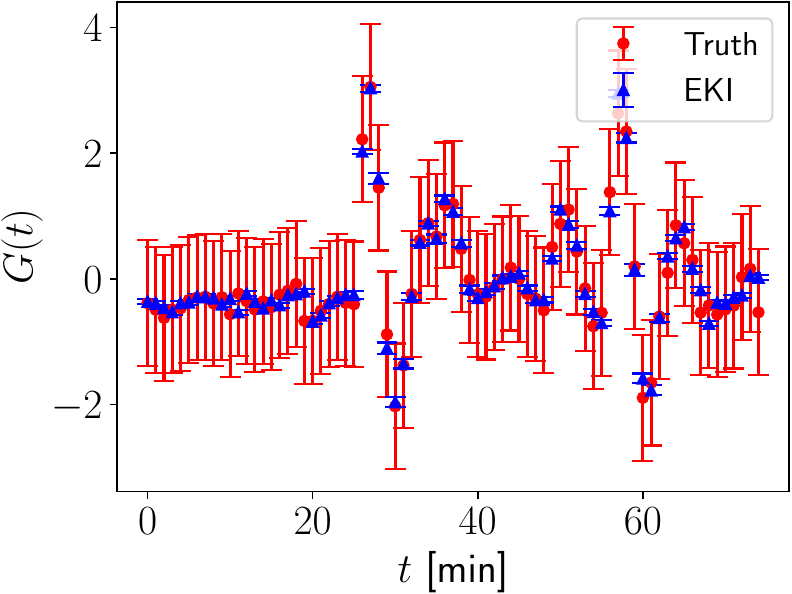} \label{fig:glucoseTraj}}
    \caption{Partial, noisy, transient timeseries data were used to calibrate model error. Results in (a-b) show the trained neural network error model for two specific input scenarios and (c) shows next-step predictions of $G$ from sequential data assimilation using the trained model. Figures (a-b) plot the model $\delta$ (a function of all 6 state variables) for 5 fixed states while varying $I_p$ and $G$, respectively.}
  \label{fig:glucoseDA}
\end{figure}

\begin{figure}[!htbp]
  \centering
  \subfloat[Trained model: $\delta$ vs $I_p$]{\includegraphics[height=0.225\textwidth, trim={1.5cm 1cm 1.5cm 2cm},clip]{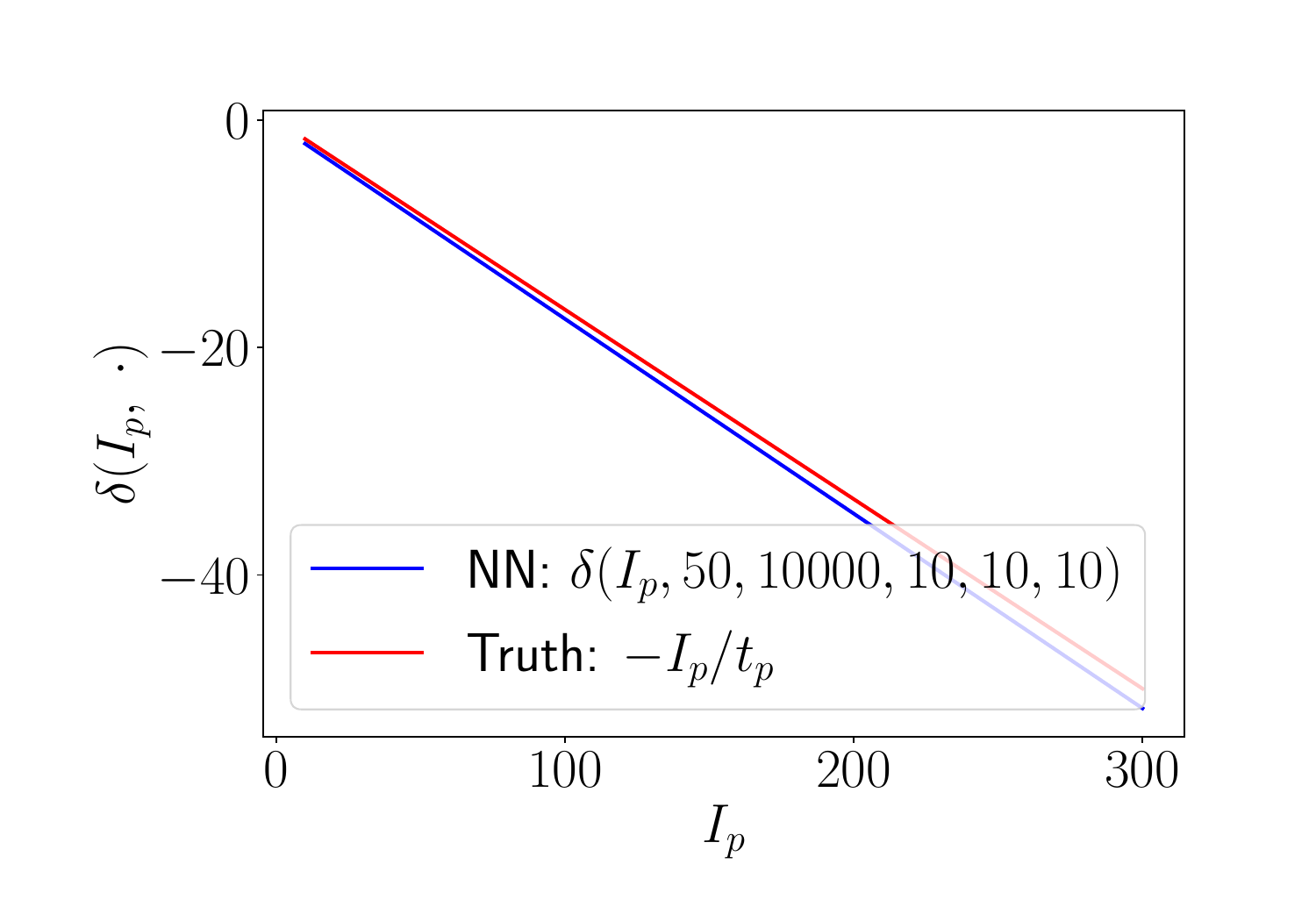}\label{fig:glucoseIp_sEKI}}
  \subfloat[Trained model: $\delta$ vs $G$]{\includegraphics[height=0.225\textwidth, trim={1.5cm 1cm 1.5cm 2cm},clip]{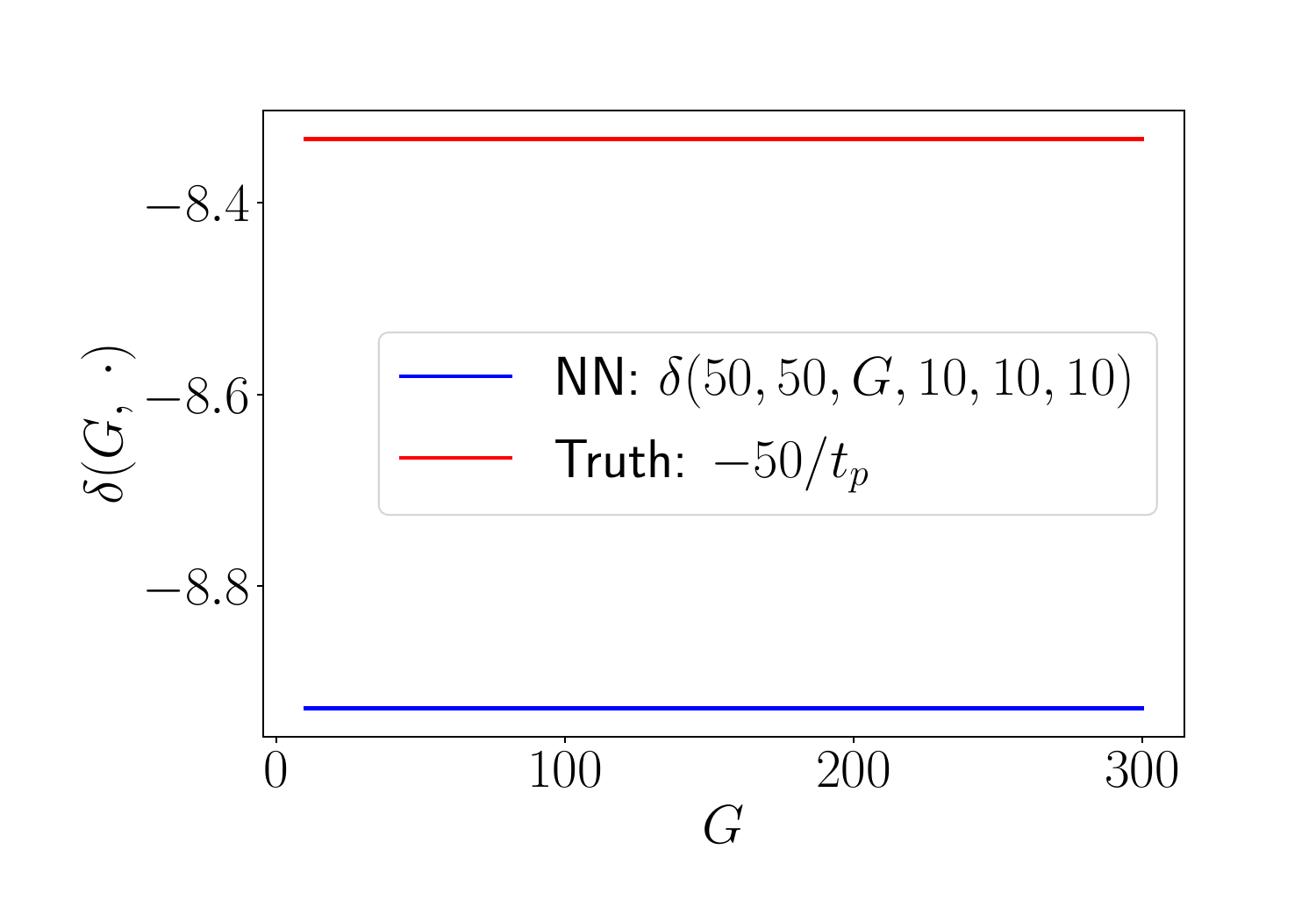} \label{fig:glucoseG_sEKI}}
\subfloat[Inferred parameter $\theta^{(\mathrm{NN})}_{20}$]{\includegraphics[height=0.225\textwidth, trim={0.8cm 0.8cm 0.7cm 0.7cm},clip]{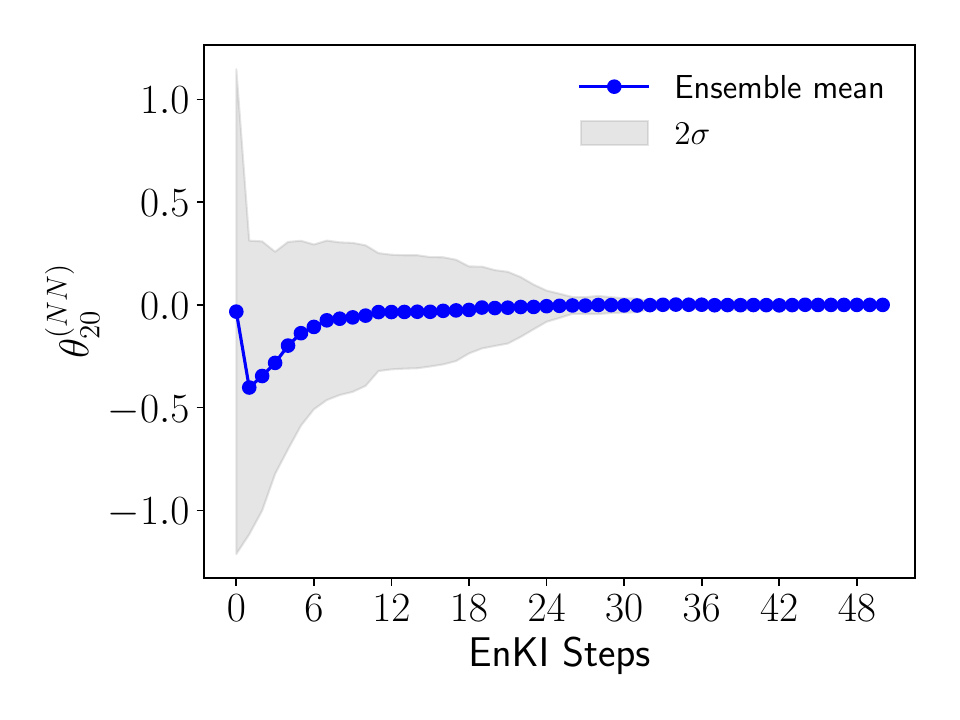} \label{fig:theta_sEKI}}
    \caption{Partial, noisy, transient timeseries data were used to calibrate model error with sparse EKI. Results in (a-b) show the trained neural network error model for two specific input scenarios and (c) shows an example of how sparse EKI drives initially non-zero error model coefficients to zero. Figures (a-b) plot the model $\delta$ (a function of all 6 state variables) for 5 fixed states while varying $I_p$ and $G$, respectively.}
  \label{fig:glucoseDA_sEKI}
\end{figure}

Through the presented numerical experiments in this section, we find that:
\begin{enumerate}
    \item Model corrections can be learned for non-ergodic, non-stationary systems from noisy, partial observations of transient dynamics by applying DA as a sub-routine within the model inference pipeline (Figure~\ref{fig:glucoseDA}).
    \item Key structural components of model error terms can be identified despite being unobserved (Figure~\ref{fig:glucoseIp}).
    \item Learned model error structure can exhibit inaccuracies outside of the observed data distribution and when models are insensitive to these inaccuracies (Figure~\ref{fig:glucoseG}).
    \item Appropriate choice of regularization can improve the inference of otherwise un-identifiable model error structure (Figure~\ref{fig:glucoseDA_sEKI}).
\end{enumerate}

Figure~\ref{fig:glucoseDA} presents results from a trained model obtained by minimizing \eqref{eq:EKI_3dvar} via EKI. Figure~\ref{fig:glucoseTraj} shows the transient trajectory data used to calibrate the model error, along with the obtained next-step predictions, which are seen to be highly accurate.
Figure~\ref{fig:glucoseIp} shows that the neural network captures the true missing linear term $-I_p / t_p$ quite
accurately for $I_p \in [0,110]$, but it is inaccurate for larger values of $I_p$. While only $G$ is observed (not $I_p$), we note that indeed $I_p(t) < 110$ for solutions of the true model. In other words, under the true model, the error term is only identifiable within the domain $[0,110]$. Figure~\ref{fig:glucoseG} shows the neural network as a function of an inactive input $G$ (an input upon
which the true missing term does not depend); instead, it learns a linear function of $G$ with a small slope.
Other results (not shown) indicate that the neural network learned substantial dependencies on inactive variables. This suggests that sparsity constraints may be useful in these experiments.

Figure~\ref{fig:glucoseDA_sEKI} presents results from a trained model obtained by minimizing \eqref{eq:EKI_3dvar} via sparse EKI~\cite{schneider2020ensemble}. Compared to standard EKI, sparse EKI incorporates sparsity into the coefficients of the trained model via $\ell_1$-constrained optimization. It can be seen in Fig.~\ref{fig:glucoseDA_sEKI}a that the neural network captures the linear pattern of the true term $-I_p / t_p$ and achieves overall better agreement than the results in Fig.~\ref{fig:glucoseDA}a. In addition, Fig.~\ref{fig:glucoseDA_sEKI}b shows that the neural network learns to correctly ignore $G$ in all plotted scenarios. An example of sparse EKI driving initially non-zero error model coefficients to zero is presented in Fig.~\ref{fig:glucoseDA_sEKI}c, for which the standard EKI would lead to larger variability with even more EKI iterations. The sparse EKI also drives some redundant coefficients of the error model to values very close to zero.

This example highlights to importance of sparsity constraints in learning error models, to avoid learning nonzero error terms that may otherwise arise because the error has zero or small projection on the output data of the model.

\section{Conclusions}
\label{sec:conclusions}
Complex nonlinear dynamics and/or a large number of degrees of freedom are present in many systems, for example, in physiological models of the human body and turbulent flows around an airplane or in the wake of wind turbines. Typically, closure models are needed if some dynamically important degrees of freedom cannot be resolved by direct numerical simulations. Without a clear scale separation between resolved and unresolved degrees of freedom, most existing models, which are, semi-empirical, deterministic and local and are calibrated with only limited amounts of data, are not sophisticated enough to capture the true dynamics of the system. Incorporating models of structural errors that are informed by data can be an effective way to build on domain knowledge while using data more extensively. 

We have summarized some key aspects of learning structural error models from data, including the construction of error models and their calibration. In doing so, we have provided guidelines about how to learn error models for complex dynamical systems, ranging from key insights about the incorporation of sparsity constraints (``do no harm'') and physical constraints, to advanced aspects such as the combined use of direct and indirect data and the merit of using non-local/stochastic error models. By addressing these varied aspects in a systematic manner, our goal has been to inspire further applied, methodological  and theoretical research in this area.

\vspace{0.1in}
\paragraph{Acknowledgments}
TS, AMS, and JW were supported by Schmidt Sciences, LLC. 
AMS is also supported by a Department of Defense Vannevar Bush 
Faculty Fellowship, and by the SciAI Center, funded by the Office of Naval Research 
(ONR), under Grant Number N00014-23-1-2729. 
MEL was supported by the National Science Foundation Graduate Research Fellowship under Grant No. DGE-1745301 and by funding from the Eric and Wendy Schmidt Center at the Broad Institute of MIT and Harvard.

All codes are available at: \\ \url{https://github.com/jinlong83/Learning-Structural-Errors.git}.


\appendix
\section{Ensemble Kalman inversion}
\label{sec:eki}

The use of ensemble Kalman based methods for parameter calibration
and the solution of inverse problems, and history of this subject, is overviewed
in [Section 4]\cite{calvello2022ensemble}. To be concrete we will concentrate on
a particular variant of the methodology, sometimes termed
Ensemble Kalman inversion (EKI). This is a specific ensemble-based, gradient-free optimization scheme that was proposed and studied in~\cite{iglesias2013ensemble}; we emphasize that other ensemble Kalman based methods share the core desirable
attributes of EKI, namely that it is derivative-free, is effective with relatively
few evaluations of the forward model $\cG$ and is robust to the presence of noise
in the evaluations of $\cG.$

The core task of EKI is equivalent to a quadratic optimization problem, which facilitates adding linear equality and inequality constraints~\cite{albers2019ensemble}. To explain the details of EKI, we first introduce a new variable $w=\cG(\theta)$ and variables $v$ and $g(v)$:
\begin{equation}
\begin{aligned}
v&=(\theta,w)^\top, \\
g(v)&=\left(\theta,\cG(\theta)\right)^\top.
\end{aligned}
\end{equation}
Using these variables we formulate the following noisily observed dynamical system:
\begin{equation}
\begin{aligned}
v_{m+1}&=g(v_m) \\
y_{m+1}&=Hv_{m+1}+\eta_{m+1}.
\end{aligned}
\end{equation}
Here $H=[0,I], H^\perp=[I,0]$, and hence $Hv=w, H^\perp v=\theta.$ 
In this setting, $\{v_m\}$ is the state and $\{y_m\}$ are the
data. The objective is to estimate $H^\perp v_m=\theta_m$ from $\{y_\ell\}_{\ell=1}^{m}$ and to do so iteratively with respect to $m$. In practice we only have one data point
$y$ and not a sequence $y_m$; we address this issue in what follows below.

The EKI methodology creates an ensemble  $\{v_{m}^{(j)}\}_{j=1}^J$ defined iteratively  in $m$ as follows:
\begin{equation}
\label{eq:optim-EKI}
\begin{aligned}
    L_m^{(j)}(v):=&\frac12\big|y^{(j)}_{m+1}-Hv\big|^2_{\Gamma}+\frac12\big|v-g\bigl({v}_{m}^{(j)}\bigr)\big|^2_{C^{gg}_{m}},\\
    v_{m+1}^{(j)}=&\argmin_v L_m^{(j)}(v).
\end{aligned}
\end{equation}
The matrix $C^{gg}$ is the  empirical covariance of $\{g(v_m^{(j)})\}_{j=1}^J$. The data $y^{(j)}_{m+1}$ is either fixed
so that $y^{(j)}_{m+1}\equiv y$
or created by adding random draws to $y$ from the distribution of
the $\eta$, independently for all $m$ and $j$.
At each step, $m$ ensemble parameter estimates 
indexed by $j=1,\cdots, J$
are found from $\theta_{m}^{(j)}=H^\perp v_{m}^{(j)}.$ 

Using the fact that $v=(\theta,w)^T$, the minimizer 
$v_{m+1}^{(j)}$ in \eqref{eq:optim-EKI} 
decouples to give the update formula
\begin{equation}
\label{eq:EKI}
\theta_{m+1}^{(j)}=\theta_{m}^{(j)}+C_m^{\theta \cG}\left(C_m^{\cG\cG}+\Gamma \right)^{-1}\left(y_{m+1}^{(j)}-\cG(\theta_m^{(j)}) \right);
\end{equation}
here the matrix $C_m^{\cG\cG}$ is the empirical covariance of $\{\cG(\theta_m^{(j)})\}_{j=1}^J$,
while matrix $C_m^{\theta \cG}$ is the empirical cross-covariance of 
$\{\theta_m^{(j)}\}_{j=1}^J$ with $\{\cG(\theta_m^{(j)})\}_{j=1}^J$.

To impose sparsity on the solution of $\theta$ from EKI, we solve the following constrained optimization problem after each EKI update step:
\begin{equation}
\label{eq:optim-EKI-lasso}
\begin{aligned}
    L_m^{(j)}(v,\lambda):=&\frac12\big|y^{(j)}_{m+1}-Hv\big|^2_{\Gamma}+\frac12\big|v-g\bigl({v}_{m}^{(j)}\bigr)\big|^2_{C^{gg}_{m}},\\
    v_{m+1}^{(j)}=&\argmin_{v \in \cV} L_m^{(j)}(v),
\end{aligned}
\end{equation}
where
\begin{equation}
\label{eq:cset1}
\cV=\{v: |H^\perp v|_{\ell_1} \le \gamma\}.
\end{equation}
We also employ the thresholding function $\mathcal{T}$ on vectors defined by
 \begin{equation}
 \label{eq:threshold}
    \mathcal{T}(\theta_i)=
    \begin{cases}
      0, & \text{if}\ |\theta_i|<\sqrt{2\lambda} \\
      \theta_i, & \text{otherwise},
    \end{cases}
  \end{equation}
to threshold those $\theta_i$ with values close to zero,
after having solved the constrained optimization problem in \eqref{eq:optim-EKI-lasso}. Such a thresholding step after the $\ell_1$-constrained optimization in \eqref{eq:optim-EKI-lasso} is equivalent to adding $\ell_0$ constraint. More details about imposing sparsity into EKI can be found in~\cite{schneider2020ensemble}.

For the multiscale Lorenz 96 example with $c=10$ and the single scale Lorenz 96 example, 100 ensembles are used and the number of EKI iterations is 20. The multiscale Lorenz 96 example with $c=3$ is more challenging to train, for which 200 ensembles are used and the number of EKI iterations is 30. The single-scale Lorenz 96 example uses 200 ensembles and the number of EKI iterations is 10. The ultradian model example is not chaotic while the convergence of standard EKI is relatively slow, for which 50 ensembles are used and the number of EKI iterations is 50. The noise level $\eta$ is estimated by running ensemble simulations of the true system and calculating the covariance of the data $y$. In practice, only the diagonal of the estimated covariance matrix is kept.

\section{Application to a Glucose-Insulin Model}
\label{sec:appendix_ult}

The functional forms of the parameterized processes in Eq.~\eqref{eq:ultradian} are as follows: 
    \allowdisplaybreaks
	\begin{align}
	    f_1(G)&=\frac{R_m}{1+\exp\left(-\frac{G}{V_g c_1}+a_1\right)} & \text{(rate of insulin production)}\label{UM8}\\
	    f_2(G)&=U_1\left(1-\exp\left(-\frac{G}{C_2V_g}\right)\right) & \text{(insulin-independent glucose utilization)}\label{UM9}\\
	    f_3(I_i)&=\frac{1}{C_3V_g}\left(U_0+\frac{U_m-U_0}{1+(\kappa I_i)^{-\beta}}\right) & \text{(insulin-dependent glucose utilization)}\label{UM10}\\
	   f_4(h_3)&=\frac{R_g}{1+\exp\left(\alpha\left(\frac{h_3}{C_5V_p}-1\right)\right)} & \text{delayed insulin-dependent glucose utilization}\label{UM11}\\
	    \kappa&=\frac{1}{C_4}\left(\frac{1}{V_i}-\frac{1}{Et_i}\right).
	\end{align}

The uptake of carbohydates is modelled by the function
	\begin{align}
	    m_G(t)=\sum_{j=1}^{N(t)} \frac{m_j k}{60 \ \text{min}} \exp(k(t_j-t)), \ \ N(t)=\#\{t_j<t\}\label{UM7}
	\end{align}
in which $N$ meals occur at times $\{t_j\}_{j=1}^N$ (in minutes), with carbohydrate composition $\{m_j\}_{j=1}^N$ (note that these typically differ from observation times).

\end{document}